\documentclass[singlecolumn,floatfix]{revtex4-2}
\usepackage{amsfonts}
\usepackage{amsmath}
\usepackage{amssymb}
\usepackage{amsthm}
\usepackage{subcaption}
\usepackage{array}
\usepackage{epsfig}
\usepackage{rotating,graphicx}
\usepackage{stix}
\usepackage{bm}%
\usepackage{verbatim}
\usepackage{float}
\usepackage{tensor}
\usepackage{xcolor}
\usepackage{picinpar}
\usepackage{tikz}
\usetikzlibrary{arrows,shapes,positioning}
\usetikzlibrary{decorations.markings}
\usetikzlibrary{arrows.meta,decorations.pathmorphing,backgrounds,fit,petri}
\usetikzlibrary{calc}
\tikzstyle arrowstyle=[scale=1]
\tikzstyle directed=[postaction={decorate,decoration={markings,
    mark=at position .7 with {\arrow[arrowstyle]{stealth}}}}]
\tikzstyle reverse directed=[postaction={decorate,decoration={markings,
    mark=at position .65 with {\arrowreversed[arrowstyle]{stealth};}}}]
\usepackage{textgreek}
\usepackage{tikz-cd}
\usepackage{multirow}
\usepackage[colorlinks=true,bookmarks=false,citecolor=blue,linkcolor=blue,hyperfootnotes=true,urlcolor=blue,linktocpage]{hyperref}
\tikzcdset{
arrow style=tikz,
diagrams={>={Straight Barb[scale=0.8]}}
}
\theoremstyle{plain}
\usepackage{amsthm}

\theoremstyle{definition}

\newtheorem{exatitle}{Example}

{\begin{exatitle} \label{#2} #1 \end{exatitle}}%
{\hfill $\Box$ \\}
\newlength{\figurewidth}
\newcommand{\ket}[1]{| #1 \rangle}
\newcommand{\bra}[1]{\langle #1 |}
\newcommand{\braket}[2]{\langle #1 | #2 \rangle}

\newcommand{\pket}[1]{[ #1 ]}

\newcommand{\td}{\text{d}}
\newcommand{\Tr}{\text{Tr}}

\newcommand{\eg}{\hbox{\em e.g.{}}}
\newcommand{\etc}{\hbox{\em etc.{}}}
\newcommand{\ie}{\hbox{\em i.e.{}}}
\newcommand{\wrt}{\hbox{w.r.t.{}}}

\newcommand{\rhs}{\hbox{r.h.s.{}}}

\newcommand{\ip}[2]{\left \langle #1, #2 \right \rangle}

\newcommand{\sjs}[6]{\left\{ \begin{array}{c c c}
#1 & #2 & #3
\\
#4 & #5 & #6
\end{array}
\right\} }

\makeatletter
\g@addto@macro\bfseries{\boldmath}
\makeatother

\usepackage{bm}

\begin{document}

\title{Speed excess and total acceleration: a kinematical approach to entanglement}

\author{C.{} Chryssomalakos}
\email{chryss@nucleares.unam.mx}
\affiliation{Instituto de Ciencias Nucleares \\
	Universidad Nacional Aut\'onoma de M\'exico\\
	PO Box 70-543, 04510, CDMX, M\'exico}
	
\author{A.{} G.{} Flores-Delgado}
\email{ana.flores@correo.nucleares.unam.mx}
\affiliation{Instituto de Ciencias Nucleares \\
	Universidad Nacional Aut\'onoma de M\'exico\\
	PO Box 70-543, 04510, CDMX, M\'exico}	
	
\author{E.{} Guzm\'an-Gonz\'alez}
\email{edgar.guzman@hainanu.edu.cn}
\affiliation{School of Physics and Optoelectronic Engineering \\
	 Hainan University\\
Haikou 570228, China}

\author{L.{} Hanotel}
\email{khanotel@hse.ru}	
\affiliation{Tikhonov Moscow Institute of Electronics and Mathematics\\
HSE University\\
Tallinskaya ul.{} 34, 123592, Moscow, Russia}

\author{E.{} Serrano-Ens\'astiga}
\email{ed.ensastiga@uliege.be}
\affiliation{Institut de Physique Nucléaire, Atomique et de Spectroscopie, CESAM
\\ Universit\'e de Liège
\\ B-4000 Liège, Belgium
}

\begin{abstract}
	\noindent
We show that the concept of total variance of a spin state, defined as the average of the variances of spin projection measurements along three orthogonal axes, also gives the rotational speed of the state in projective space, averaged over all rotation axes. We compute the addition law, under system composition, for this quantity and find that, in the case of separable states, it is of simple pythagorean form. In the presence of entanglement, we find that the composite state ``rotates faster than its parts'', thus unveiling a kinematical  origin for the correlation of total variance with entanglement. We analyze a similar definition for the acceleration of a state under rotations, for both pure and mixed states, and probe numerically its relation with a wide array of entanglement related measures.
\end{abstract}
\date{\today}
\maketitle

\tableofcontents
\newsavebox{\Potb}
\section{Introduction}
\label{Intro}
Quantum entanglement has been the focus of intense activity, both experimental and theoretical,  for several decades now, owing to its intrinsic appeal as the epitome of quantum counterintuitiveness, on the one hand, as well as its relevance in quantum technology applications, on the other~\cite{Ben.Zyc:17,Nie.Chu:11,Kus.Zyc:01}. Central to its very definition, is the division of a physical system in subsystems --- subsequent to that, quantifying entanglement, and related concepts, is a surprisingly multifaceted affair, that becomes increasingly convoluted as the number of subsystems considered grows. 

Among the plethora of available measures of entanglement (see, \eg,~\cite{Bru:02}), one that stands out is Klyachko's ``total variance''~\cite{Kly:07} --- applied to spin states (which is the case we focus on here)  it averages the variance of spin projection measurements over three orthogonal axes (see~(\ref{totvardef}) for the exact formula). In the author's words ``\ldots It measures the total level of quantum fluctuations of the system\ldots''.  There are many instances where this quantity appears naturally, for example, it is easily shown that a spin state is coherent (in many respects ``most classical'') if and only if it minimizes total variance~\cite{Del.Fox:77}, a desirable property for an entanglement measure, since coherent states, viewed as symmetrized states of spin-1/2 subsystems, are separable. At the other extreme, 1-anticoherent spin states~\cite{Zim:06}, which have vanishing spin expectation value, maximize total variance, in accord with their (informal) status as the ``most quantum'' states. Thus, Klyachko's proposal seems to pass some basic consistency checks, but a lingering (for the authors) question is ``what has spin variance to do with entanglement?'' --- this is the starting point of our present work.

Other authors before us have focused their attention to total variance, finding, \eg, an intriguing relation between its critical sets and SLOCC classes~\cite{Saw.Osz.Kus:12}, a result that further motivated us to understand the definition at an intuitive level. What we find can be summarized as follows:
\begin{itemize}
\item
The total variance of a state is its average squared speed (in projective space, using the Fubini-Study metric) under rotations.
\item
In a composite system, entangled states attain higher speeds, on the average, than separable ones when rotated. 
\item
``Addition laws'' for the average squared rotational speed, and similar quantities, involving higher time derivatives, provide a kinematical point of view on entanglement that is worth exploring.
\end{itemize}
Guided by these initial findings, we propose a ``speed excess'' measure of entanglement: if $|v|^2$ denotes the average squared rotational speed of a bipartite state, and $|v_1|^2$, $|v_2|^2$ the analogous quantity for its subsystems, then the extent to which the separable state pythagorean addition law $|v|^2=|v_1|^2+|v_2|^2$ is violated can be taken as a measure of entanglement. Following further this line of reasoning, we look for an addition law of average squared acceleration, due to rotations, and find that, in the separable case, the subsystems contribute to the acceleration of the bipartite state not only through their own acceleration, but also through their speed. For entangled states, where the reduced states of the subsystems are mixed, the problem becomes more complicated, as the Fubini-Study metric gets replaced by the Bures metric. We resort to numerical methods in order to explore correlations of these kinematical quantities with several other physically relevant measures.

Entanglement has been related to ``speed of evolution'' before, starting with the study of energy-time uncertainty relations~\cite{Man.Tam:45}, the subsequent realization that energy uncertainty relates to speed of evolution~\cite{Aha.Ana:90}, followed by inquiries into the maximum attainable speed~\cite{Mar.Lev:98,Lev.Tof:09,Bat.Cas.Pla.Pla:05,Bor.Zan.Pla.Cas.Pla:08}, and the role of entanglement in achieving it~\cite{Gio.Llo.Mac:03,Gio.Llo.Mac:03a}. Our present contribution complements the above by focusing on the average rotational speed, giving its precise quantitative relation with entanglement, in the form of an addition law, and generalizing these concepts to higher covariant derivatives of the curve traced in quantum state space.

The structure of the paper is as follows: after some standard background material  in Sect.~\ref{MathPre}, we interpret, in Sect.~\ref{TVEaSE}, the total variance of a state as average rotational speed squared, and generalize the concept for mixed states. In Sect.~\ref{OtaociP} we introduce the total (average) acceleration (squared) of a pure state and give general expressions for it for any spin. Sect.~\ref{AAL}  
derives the acceleration addition law for pure separable states and then treats the mixed state case. In Sect.~\ref{KQPaQC} we present an extensive collection of plots exploring the correlation among the newly defined quantities and well known related measures like linear and von Neumann entropy, concurrence, negativity, \etc. Some final remarks appear in Sect.~\ref{Conclusions}.

\section{Mathematical Preliminaries}
\label{MathPre}
\subsection{The projective space $\mathbb{P}$}
We use the notation and conventions in~\cite{Chr.etal:23}, which we briefly review here. Quantum states of a spin-$s$ system are represented by a vector (ket) $\ket{\psi} \in \mathcal{H} \equiv \mathbb{C}^{n+1}$, with $n=2s$. States that differ by (complex) rescaling are in a certain physical sense equivalent, there is therefore a natural projection $\Pi$ to the equivalence class $\pket{\psi} \in \mathbb{C}P^{n}$,
\begin{equation}
\label{piproj}
\Pi \colon \mathcal{H} \rightarrow \mathbb{C}P^n, \quad \ket{\psi}
=
(\psi^0,\psi^1,\ldots,\psi^n)^T \mapsto \pket{\psi}=(z^1,\ldots,z^n)
\, ,
\end{equation}
with $z^i=\psi^i/\psi^0$, together with their complex conjugates  $\bar{z^i} \equiv w^i$, serving as coordinates in the $U_0$ chart, where $\psi^0 \neq 0$ --- we will denote them collectively by $z^A$, with $A$ ranging over $\{1,\ldots,n,\bar{1},\ldots,\bar{n}\}$.  $\mathbb{C}P^n$, in its turn, may be embedded into $\mathfrak{u}(n+1)$ as the $U(n+1)$-adjoint orbit of the density matrix $\rho_0=\text{diag}(1,0,\ldots,0)$ (see, \eg,~\cite{Ban.Hur:04}), the latter living naturally in the unitary Lie algebra $\mathfrak{u}(n+1)$ (in its hermitian version preferred by physicists),
 \begin{equation}
 \label{CPn2un}
 \varphi \colon \mathbb{C}P^n \hookrightarrow \mathfrak{u}(n+1) \, ,
 \quad
 \pket{\psi} \mapsto \rho_\psi=\frac{\ket{\psi}\bra{\psi}}{\braket{\psi}{\psi}}=
 \Delta^{-1}
 \left(
 \begin{array}{cccc}
 1 & w^1 & \ldots & w^n
 \\
 z^1 & z^1 w^1 & \ldots & z^1 w^n
 \\
 \vdots & & & \vdots
 \\
 z^n & z^n w^1 & \ldots & z^n w^n
 \end{array}
 \right)
 \, ,
 \end{equation}
 with $\Delta \equiv 1+\sum_{i=1}^n z^i w^i$.  We abbreviate the image of $\mathbb{C}P^n$ under $\varphi$ by $\mathbb{P} \subset \mathfrak{u}(n+1)$. Thus, $\mathbb{P}$ is the locus of $(n+1)$-dimensional complex matrices $\rho$ satisfying
 \begin{equation}
 \label{rhoprops}
 \rho^\dagger=\rho
 \, ,
 \qquad
 \Tr \, \rho=1
 \, ,
 \qquad
 \rho^2=\rho
 \, .
 \end{equation}
 We use greek indices, ranging from 0 to $n$, for the components of  vectors and density matrices, and slightly abuse that notation writing  $(z^{\mu})=(1,z^1,\ldots,z^n)$, so that, \eg,
 \begin{equation}
 \label{rhomunu}
 \rho^{\mu \nu}=\Delta^{-1} z^\mu w^\nu
 \, .
 \end{equation}
 The coordinate basis in the tangent space $T_\rho \mathbb{P}$ is  given by  $\rho_A \equiv \partial \rho /\partial z^A$, which are \emph{not} hermitian matrices. Real tangent vectors $v$ are constrained to satisfy $v^{\bar{a}}=\overline{v^a}$, with $v^{\bar{a}}$ denoting the component of $v$ along $\partial_{w^a}$, $v=v^a \partial_{z^a}+v^{\bar{a}} \partial_{w^a} \equiv v^a \partial_a +v^{\bar{a}} \partial_{\bar{a}}$. Tangent vectors to $\mathbb{P}$, like the above $v$,  are  $(n+1)$-dimensional complex matrices satisfying the infinitesimal versions of~(\ref{rhoprops}), 
 \begin{equation}
 \label{vprops}
 v^\dagger=v
 \, ,
 \qquad
 \Tr \, v=0
 \, ,
 \qquad
 \rho v+ v \rho=v
 \, .
 \end{equation}
 The natural (Fubini-Study (FS)) metric on $\mathbb{P}$ is $\ip{v}{v'}=\frac{1}{2}\Tr(vv')$.
 In the above coordinate basis, the FS metric and its inverse have components 
\begin{equation}
\label{FSmcomp}
g_{a\bar{b}}=\frac{1}{2}\Delta^{-2}(\Delta\delta^a_{\phantom{a}b}- z^b w^a)
\, ,
\quad
g^{a\bar{b}}=2\Delta(\delta^a_{\phantom{a}b}+z^a w^b)
\, ,
\end{equation}
with $g_{\bar{b}a}=g_{a\bar{b}}$ (\ie, $g_{AB}$ is symmetric), and $g_{b\bar{a}}=\bar{g}_{a\bar{b}}$ (\ie, $(g_{a\bar{b}})$ is hermitian) and similar statements holding true for the inverse metric.
The Christoffel symbols are found to be
\begin{equation}
\label{ChrComp}
\Gamma^{c}_{\phantom{c}ab}
=
g^{c\bar{r}} \partial_a g_{b\bar{r}}
=
-\Delta^{-1}( \delta^c_{\phantom{c}b} w^a+\delta^c_{\phantom{c}a} w^b)
\, ,
\quad
\Gamma^{\bar{c}}_{\phantom{\bar{c}}\bar{a}\bar{b}}=-\Delta^{-1}( \delta^c_{\phantom{c}b} z^a+\delta^c_{\phantom{c}a} z^b)
\, ,
\end{equation}
with all mixed components vanishing, while  the non-zero Riemann tensor components are
\begin{equation}
\label{Rcomp}
R_{a\bar{b}c\bar{d}}=\frac{1}{2}(g_{a\bar{b}} g_{c\bar{d}}+g_{a\bar{d}} g_{c\bar{b}})
\, .
\end{equation}
Given a curve $\rho_t$ in $\mathbb{P}$, its velocity $v_t=\dot{\rho}_t$ is tangent to  $\mathbb{P}$ at $\rho_t$, but its second time derivative is, in general not --- the acceleration of $\rho_t$, which we define as the covariant time derivative of $v_t$, using the Levi-Civita connection of the FS metric,  can be obtained from $\ddot{\rho}_t$ by projecting it onto $T_{\rho_t}\mathbb{P}$ (see~\cite{Chr.etal:23} for details)
\begin{equation}
\label{accdef}
a_t=\nabla_t v_t =\ddot{\rho}_t^\parallel =\rho_t \ddot{\rho}_t \tilde{\rho}_t+\tilde{\rho}_t \ddot{\rho}_t \rho_t
\, ,
\end{equation}
where $\tilde{\rho}_t \equiv I-\rho_t$ (in what follows we often omit writing the subindex $t$).
\subsection{Hilbert-Schmidt space and the Bures metric}%
\label{sec:app1}
\subsubsection{Horizontality in the Hilbert-Schmidt space}%
\label{sec:HorHS}
The Hilbert-Schmidt space $\mathcal{H \! S}(\mathcal{H})$ of a Hilbert space $\mathcal{H}$ is the (also Hilbert) space of linear operators acting on $\mathcal{H}$, equipped with the hermitian inner product $\ip{A}{B}=\frac{1}{2} \Tr(A^\dagger B)$. The latter gives rise to the $\mathcal{H \! S}$ metric
\begin{equation}
\label{HSmetric}
\overline g(A,B) = \frac{1}{2} \Tr ( A^{\dagger} B +BA^{\dagger}  ).
\end{equation}
There is a natural projection $\pi \colon \mathcal{H \! S}^* \rightarrow P$ (where $\mathcal{H \! S}^*$ denotes the subset of invertible operators, and $P$ the positive cone, \ie, the space of positive (and, hence, hermitian) operators), given by
\begin{equation}
\pi(A)= A A^{\dagger}
\, ,
\qquad
\pi_{*}(\dot A)=
 \dot A A^{\dagger} + A \dot A^{\dagger}
 \, ,
\end{equation}
where $\pi_*$ denotes the corresponding pushforward. 
The unitary group acts from the right on $\mathcal{H \! S}$, $A \triangleleft U = A U$ and the orbit of $A$ under this action is the fiber of $\pi$, \ie, the locus of operators that share the same projection. The vectors tangent to the fiber are declared to be vertical --- a hermitian matrix $H$ gives rise to a vertical vector field $H^\sharp(A)\equiv-iAH$. Tangent vectors $V$ that are perpendicular to all vertical vectors are declared horizontal,
\begin{equation}
\overline g(V,  H^{\sharp}) = 0 \text{ for all hermitian $H$} \Rightarrow V \text{ horizontal}
\, .
\end{equation}
Unpacking the previous expression we get
\begin{equation}
\overline g(V,  H^{\sharp}) = 0  \Leftrightarrow \Tr (i V H A^{\dagger} - i V^{\dagger} A H) = 0
    \Leftrightarrow
\Tr  [H (A^{\dagger} V -  V^{\dagger} A)]  = 0
\, ,
\end{equation}
which, being true for all hermitian $H$, implies that $A^{\dagger} V =  V^{\dagger} A$,
that is, $A^{\dagger} V = F$ is hermitian. Solving for $V$ gives
\begin{equation}
V = (A^{-1})^{\dagger} F = (A^{-1})^{\dagger} F  A^{-1} A = G A,
\end{equation}
where $G=(A^{-1})^{\dagger} F  A^{-1}$ is also hermitian. Thus, horizontality of $V \in T_A \mathcal{H \! S^*}$ implies $V=GA$, with $G$ hermitian --- it is easily seen that the converse is also true. Given a curve $P= \pi (A)$ in the space of positive operators,
with $A$  horizontal (\ie, with its tangent vector $\dot{A}$ horizontal), we get
\begin{equation}
\label{dPGP}
\dot P = G P + P G
\, .
\end{equation}
More details can be consulted in~\cite{Uhlmann1986229} and references therein, see also~\cite{Ben.Zyc:17}.
\subsubsection{The Bures metric}
We restrict the previous discussion to operators $A$ on the sphere  $S$ defined by  $\Tr(A A^{\dagger})=1$.
The image of $S$ under $\pi$ gives the space of density matrices $\rho$, where the Bures metric is defined.
The tangent space $T_A S$ consists of all vectors $V$ orthogonal to $A$,
\begin{equation}
\Tr(V A^{\dagger}+ V^{\dagger} A)= 0 \Leftrightarrow
\overline g(V,A) = 0
\, .
\end{equation}
Given a vector $V \in T_A \mathcal{H \! S^*}$, its component orthogonal to $T_A S$ is
\begin{equation}
V^\perp = V- \frac{1}{2}\Tr(AV^{\dagger}+ V A^{\dagger}) A
\, ,
\end{equation}
so that, given a curve $A \in S$, its intrinsic acceleration $a$ can be computed by projecting $\ddot A$ onto $T_{A}(S)$, 
\begin{equation}
\label{addproj}
  \begin{split}
 \overline a = \ddot A - \frac{1}{2}\Tr(A\ddot A^{\dagger}+ \ddot A A^{\dagger}) A.
  \end{split}
\end{equation}

Finally, the Bures metric is defined by $g(\dot{\rho}_1,\dot{\rho}_2)=\overline{g}(\dot{A}_1,\dot{A}_2)$, where $A_i$ are horizontal lifts of $\rho_i$. It follows that
\begin{align}
g( \dot \rho_1, \dot \rho_2 ) 
&= 
\overline g (\dot A_{1},\dot A_{2})
\nonumber
\\
&= 
\frac{1}{2} \Tr (\dot A_{1}^{\dagger} \dot A_{2}+\dot A_{2}^{\dagger} \dot A_{1} )
\nonumber
\\
&= 
\frac{1}{2} \Tr ( A_{1}^{\dagger} G_{1} \dot A_{2}+\dot A_{2}^{\dagger} G_{1} A_{1} )
\nonumber
\\
&= 
\frac{1}{2} \Tr ( G_{1} (\dot A_{2}A_{1}^{\dagger} + A_{1}\dot A_{2}^{\dagger}) )
\nonumber
\\
&= \frac{1}{2} \Tr ( G_{1}  \dot \rho_2 )
\, ,
\label{grho12}
\end{align}
where $G_{1} A_{1}= \dot A_{1}$, and $A_1(0)=A_2(0)=A$, $\rho_1(0)=\rho_2(0)=\rho$, \ie, both curves $A_i$ emanate from $A$, and similarly for $\rho_i$. Alternatively,
\begin{equation}
g( \dot \rho_1, \dot \rho_2 )=
 \frac{1}{2} \Tr \left( G_{1}  (G_{2}\rho+ \rho G_{2}) \right)
= \frac{1}{2} \Tr \left(  \rho (G_{1}  G_{2}+ G_{2} G_{1}) \right)
\, .
  \label{eq:altgg}
\end{equation}
\section{Total Variance, Entanglement, and Speed Excess}
\label{TVEaSE}
\subsection{The many facets of total variance}
\label{Tmfotv}
\subsubsection{Total variance as a measure of quantum fluctuations}
\label{Tvoaqs}
Given a Lie group $G \subset U(n)$ which acts on $\mathbb{P}$, and a linear basis $\{e_A, \, A=1,\ldots,k\}$ of the corresponding Lie algebra $\mathfrak{g} \subset \mathfrak{u}(n)$, orthonormal \wrt{} an $\text{ad}_\mathfrak{g}$-invariant metric, the  \emph{total $\mathfrak{g}$-variance} of $\rho =\ket{\psi}\bra{\psi} \in \mathbb{P}$ (with $\ket{\psi}$ in $\mathcal{H}$) is defined in~\cite{Kly:07} (see also~\cite{Del.Fox:77}) as 
\begin{align}
\label{totvardef}
\mathbb{D}_{\mathfrak{g}}(\rho)
&=
\sum_{A=1}^k \bra{\psi} T_A^2\ket{\psi} -\bra{\psi} T_A \ket{\psi}^2
\, ,
\end{align}
where $T_A$ is the matrix representing the action of $e_A$ on $\mathcal{H}$. The obvious physical interpretation of this quantity is as a measure of ``\ldots the total level of quantum fluctuations of the system in state $\ket{\psi}$''~\cite{Kly:07}. In the case of $G=SU(2)$ and $\mathbb{P}=\mathbb{C}P^{2s}$, the first term in the \rhs{} of~(\ref{totvardef}) gives the $SU(2)$ Casimir operator in the spin-$s$ representation, so that
\begin{align}
\label{totvarsu2}
\mathbb{D}_{\mathfrak{su}(2)}(\rho)
&=
s(s+1) -\sum_{A=1}^3 \bra{\psi} S_A \ket{\psi}^2
\, ,
\end{align}
implying that the total variance is minimized by coherent states, and maximized by anticoherent ones (the latter defined by the vanishing of the spin expectation value $\bra{\psi}\mathbf{S}\ket{\psi}$~\cite{Zim:06}).
An additional,  less than obvious, physical interpretation of $\mathbb{D}_{\mathfrak{su}(2)}(\rho)$ is also put forth in \cite{Kly:07}: considering a spin-$s$ state  as a multipartite symmetric state of $2s$ spin-1/2 subsystems, the suggestion is made that its total $\mathfrak{su}(2)$-variance be considered as a measure of its entanglement. The idea has been further explored in~\cite{Saw.Osz.Kus:12}, where the critical sets of $\mathbb{D}_{\mathfrak{su}(2)}$ are used to classify all SLOCC classes of multipartite pure states. 
\subsubsection{Total variance as a measure of rotational speed}
\label{Tvaam}
We propose here an alternative characterization of the total variance, that, in turn, suggests an explanation of its relation with entanglement. When a spin-$s$ system is rotated in physical space, around an axis $\mathbf{n}$, the velocity of its quantum state in $\mathbb{P}$ is 
\begin{equation}
\label{velrotn}
v_{\mathbf{n}}=-i[\mathbf{n}\cdot \mathbf{S},\rho]
\, ,
\end{equation}
with modulus squared
\begin{align}
|v_{\mathbf{n}}|^2
&=
\frac{1}{2} \text{Tr} \, v_{\mathbf{n}}^2
\nonumber
\\
&=
-\frac{1}{2} \text{Tr} 
\left( 
\left[
\mathbf{n}\cdot \mathbf{S},\rho
\right]
\left[
\mathbf{n}\cdot \mathbf{S},\rho
\right] 
\right)
\nonumber
\\
&=
\bra{\psi} (\mathbf{n} \cdot \mathbf{S})^2 \ket{\psi} - \bra{\psi} \mathbf{n} \cdot \mathbf{S} \ket{\psi}^2
\label{velsqn}
\, .
\end{align}
Averaging over the rotation axis (with $\int_{S^2} n_i n_j=\delta_{ij}/3$) we find
\begin{equation}
\label{vnsqD}
\left \langle |v_{\mathbf{n}}|^2 \right  \rangle_{S^2} = \frac{1}{3} \mathbb{D}_{\mathfrak{su}(2)}(\rho)
\equiv \frac{1}{3} \mathbb{D}(\rho)
\, ,
\end{equation}
\ie, the total variance is proportional to the square of the rotational speed of the state, averaged over the rotation axis (from now on, we simplify the notation by dropping the $\mathfrak{su}(2)$ index). This geometric interpretation of the total variance suggests an obvious generalization to mixed states. The velocity $v_{\mathbf{n}}$ acquired by such a state $\rho$ when rotated around $\mathbf{n}$ is still given by~(\ref{velrotn}), but its modulus squared $|v_{\mathbf{n}}|^2$ entails now the Bures metric~\cite{Ben.Zyc:17}. Averaging $|v_{\mathbf{n}}|^2$ over $\mathbf{n}$ gives the total variance of the mixed state $\rho$ --- we give an example of an explicit calculation below.
\subsubsection{Total variance as a measure of entanglement: speed excess}
\label{Salae}
Aiming at connecting total variance to entanglement, we ask now what is the addition law for the square of the speed of a composite quantum system, given the square of the speeds of its subsystems? For a separable state, $\rho=\rho_1 \otimes \rho_2$, one finds easily 
\begin{equation}
\label{spadlawPytha}
|v|^2=|v_1|^2+|v_2|^2
\, ,
\end{equation}
where $v=\dot{\rho}$, $v_1=\dot{\rho}_1$, $v_2=\dot{\rho}_2$, $|v|^2=1/2 \, \text{Tr} \, v^2$ \etc --- this result is valid for an arbitrary time evolution, \ie, it is not tied to rotations.  Anticipating that the presence of entanglement will complicate things, in the form of additional, entanglement-dependent terms, in the \rhs{} of~(\ref{spadlawPytha}), we define the \emph{(squared) speed excess} $F(\rho)$ of a pure bipartite state $\rho$ as follows
\begin{equation}
\label{spexc}
F(\rho)=|v|^2-|v_1|^2-|v_2|^2
\, ,
\end{equation}
where now $\rho_1=\text{Tr}_2 \rho$, $\rho_2=\text{Tr}_1 \rho$ are the reduced density matrices, corresponding, in general, to mixed states,  and the moduli $|v_i|^2$ are computed, as mentioned above,  with the Bures metric. While the separable-state addition law~(\ref{spadlawPytha}) implies that the speed of a composite system is entirely due to the speed of its parts, the more general addition law in~(\ref{spexc}) implies that, in general, entanglement also contributes to the speed of the composite system, leaving open the possibility \eg, that the latter may move even when its parts are ``at rest''. Consider, for example, the bipartite  spin-1/2  symmetric state $\ket{\Psi_t}$ with Majorana constellation given by two antipodal stars on the equator, rotating around the $z$-axis with unit angular velocity, expressed in the $(\ket{++},\ket{+-},\ket{-+},\ket{--})$-basis,
\begin{equation}
\label{Psipm}
\ket{\Psi_t}=\frac{1}{\sqrt{2}} \left( 1,0,0,-e^{i2t} \right)
\, .
\end{equation}
 This composite state certainly has nonzero speed in projective space (essentially the spin-1 state space $\mathbb{C}P^2$) --- one easily computes that $|v|^2=|\dot{\rho}_{\ket{\Psi}}|^2=1$, while the reduced density matrices are those of the maximally mixed state, with vanishing speeds, $|v_1|^2=|v_2|^2=0$ --- in this case, the entire speed of the composite state is due to entanglement, and $F(\rho_{\ket{\Psi_t}})=1$.

Specifying~(\ref{spexc}) to the case of symmetric states of two spin-1/2 systems, where the reduced density matrices are equal among themselves, $\rho_1=\rho_2$, we get
\begin{equation}
\label{spexc2s}
F(\rho)=|v|^2-2|v_1|^2
\, ,
\end{equation}
with 
\begin{equation}
\label{vardefs1}
\rho_1=\frac{1}{2} (I+ \mathbf{r} \cdot \boldsymbol{\sigma})
\, ,
\quad
\mathbf{r}=(x,y,z)
\, ,
\quad
v_1=\dot{\rho}_1
\, , 
\quad
|v_1|^2=g_{AB}(\rho_1)v_{1 \, A} v_{1 \, B}
\, ,
\quad
v_{1 \, A} =\text{Tr} \, (v_1 \sigma_A)
\, ,
\end{equation}
and the Bures metric being given by (see the metric in (9.50) of~\cite{Ben.Zyc:17}, which should be pulled back onto the unit-trace submanifold)
\begin{equation}
\label{gBdef}
g(\rho_1)=
\frac{1}{4(1-x^2-y^2-z^2)}
\left(
\begin{array}{ccc}
1-y^2-z^2 & x y & x z
\\
x y & 1-x^2-z^2 & y z
\\
x z & y z & 1-x^2-y^2
\end{array}
\right)
\, .
\end{equation}
When the time evolution of $\rho$ is due to rotation around $\mathbf{n}$, we get (compare with~(\ref{velrotn}))
\begin{equation}
\label{vncomp}
v_{\mathbf{n}}=-\frac{i}{2} \left[ 
\mathbf{n} \cdot (\boldsymbol{\sigma} \otimes I 
+
I \otimes \boldsymbol{\sigma}),\rho 
\right]
\, ,
\quad
v_{1 \, \mathbf{n}}=-\frac{i}{2}
\left[ 
\mathbf{n} \cdot \boldsymbol{\sigma},\rho_1
\right]
\, ,
\end{equation}
from which we can calculate $F_{\mathbf{n}}(\rho)=|v_{\mathbf{n}}|^2-2|v_{1 \, \mathbf{n}}|^2$. Averaging this over $\mathbf{n}$ we finally get the \emph{total (rotational) speed excess}
\begin{align}
\mathbb{F}(\rho) &\equiv \left \langle F_{\mathbf{n}}(\rho) \right \rangle_{S^2}
\nonumber
\\
&=
\left \langle |v_{\mathbf{n}}|^2 \right \rangle_{S^2}-2 \left \langle |v_{1 \, \mathbf{n}}|^2 \right \rangle_{S^2}
\, .
\label{avrse}
\end{align}

On the other hand, a straightforward calculation, starting from the definition~(\ref{avrse}), and using (\ref{vardefs1}), (\ref{gBdef}), and (\ref{vncomp}),  shows that
\begin{align}
\mathbb{D}(\rho) +
\mathbb{D}(\rho_1) +
\mathbb{D}(\rho_2) 
&=
3 \left( \left \langle |v_{\mathbf{n}}|^2 \right \rangle_{S^2}+2 \left \langle |v_{1 \, \mathbf{n}}|^2 \right \rangle_{S^2} \right)
\nonumber
\\
&=
|v_{\mathbf{x}}|^2 + |v_{\mathbf{y}}|^2 + |v_{\mathbf{z}}|^2
+2 \left( |v_{1 \, \mathbf{x}}|^2+|v_{1 \, \mathbf{y}}|^2 + |v_{1 \, \mathbf{z}}|^2 \right)
\nonumber
\\
&=
2
\, ,
\label{v2v12rel}
\end{align}
which, at first, might look like a counterintuitive result (at least it did to the authors): it says that the bigger the total variance of the composite state $\rho$ is, the smaller the total variances of the states of the subsystems have to be, and \emph{vice versa}. A moment's thought though reveals that this result exactly encodes, in a precise quantitative manner, the fact that total variance is a measure of entanglement --- we proceed to explain this statement in some detail: we start with expressing the Bures metric in spherical polar coordinates $(r,\theta,\phi)$, where it acquires a diagonal form,
\begin{equation}
\label{gBpolar}
g^{\text{polar}}(\rho_1)=\frac{1}{4}\text{diag} \left( \frac{1}{1-r^2},r^2,r^2 \sin^2\theta \right)
\, ,
\end{equation}
where $r=\sqrt{x^2+y^2+z^2}$, $\cos\theta=z/r$. Note that this differs from the euclidean metric $\text{diag}(1,r^2,r^2\sin^2\theta)/4$, which is the trace metric~(\ref{velsqn}), only in the radial direction. But $SU(2)$ transformations of $\rho$ result in rotations of $\mathbf{r}$, the corresponding velocity having no radial component. Thus, the speed, due to rotations,  of the reduced state $\rho_1$, calculated with the Bures metric,  is just the euclidean speed of the tip of $\mathbf{r}$, which, for a given rotation, only depends on the length $r$ of the vector. As a result, the higher the total variance of the bipartite state $\rho$ is, the smaller $r$ has to be (because of~(\ref{v2v12rel})), the more mixed the reduced state $\rho_1$ is, and, hence, the more entangled $\rho$ turns out to be.

Having clarified this point, we return to our calculation of the total speed excess. Using~(\ref{v2v12rel}), together with~(\ref{vnsqD}), (\ref{avrse}), we arrive at
\begin{equation}
\label{DFrel}
\mathbb{F}(\rho)
=
\frac{2}{3} \left (\mathbb{D}(\rho)-1 \right)
\, ,
\end{equation}
in other words, total speed excess and total variance are functions on the symmetric subspace of the 2-qubit state space (\ie, on the spin-1 projective space $\mathbb{C}P^2$) that are related by a simple affine transformation. The total speed excess is minimum on the coherent states (equal to zero) and maximum on the anticoherent ones (equal to 2/3). Since $\mathbb{F}$ in~(\ref{DFrel}) is non-negative, we conclude that, in the case of bipartite spin-1/2 symmetric pure states,  \emph{entanglement increases rotational speed}.
\subsection{Total variance  for mixed states}
\label{Tvfms}
We inquire about the extension of the total variance concept to bipartite mixed states. Restricting our analysis, as in the pure state case above, to the symmetric sector of a two-qubit system, a general density matrix can be parametrized as follows
\begin{equation}
\label{rhomixedpara}
\rho=\frac{1}{4} I 
+ \mathbf{n} \cdot \boldsymbol{\Sigma}
+ \frac{1}{4}\sum_{A=1}^3 t_{AA} \Sigma_{AA}
+ \frac{1}{8} \sum_{B>A=1}^3 t_{AB} \Sigma_{AB}
\, ,
\end{equation}
where
\begin{equation}
\label{Sigmadefs}
\boldsymbol{\Sigma}=(\Sigma_1,\Sigma_2,\Sigma_3)
\, ,
\quad
\Sigma_A=\sigma_A \otimes I+I \otimes \sigma_A
\, ,
\quad
\Sigma_{AB}=\sigma_A \otimes \sigma_B +\sigma_B \otimes \sigma_A
\, ,
\end{equation}
with appropriate conditions on $\mathbf{n}$, $t_{ij}$ guaranteeing non-negative eigenvalues for $\rho$ (see, \eg,~\cite{Gam:16}) --- note that for vanishing $\mathbf{n}$, $t_{ij}$ all eigenvalues of $\rho$ are equal to 1/4, so we can be sure that for sufficiently small values of these parameters all eigenvalues of $\rho$ will be positive. The modulus squared, according to the Bures metric,  of a tangent vector $X$ at $\rho$ is given by 
\begin{equation}
\label{gBXX}
|X|^2_{\text{B}}=g_{\text{B}}(X,X)_\rho=\frac{1}{2} \text{Tr} \left( GX \right)
\, ,
\end{equation}
where the hermitian matrix $G$ is uniquely determined (for positive $\rho$) by the relation $X=\rho \, G +G \rho$. We now map an $n \times n$ matrix $A$ to an $n^2$-dimensional vector $\ket{A}=(A_{11}, A_{12},\ldots,A_{nn})^T$ --- it is easily checked that, in this notation, $(A \otimes B^T)\ket{C}=\ket{ACB}$ (where $(A \otimes B^T)_{ij,rs}=A_{ir}B_{sj}$), so that the above relation for $G$ becomes
\begin{equation}
\label{Gdefeq}
\ket{X}= R \ket{G}
\, ,
\quad
R \equiv \rho \otimes I + I \otimes \rho^T
\, .
\end{equation}
It is easily seen that the eigenvalues of $R$ are $\{ \lambda_\alpha+\lambda_\beta\}$, where $\{\lambda_\alpha \}$ are those of $\rho$, so that $R$ is invertible if $\rho$ is positive, and~(\ref{Gdefeq}) gives $\ket{G}=R^{-1} \ket{X}$, resulting in
\begin{align}
g_{\text{B}}(X,X)_\rho
&=
\frac{1}{2} \sum_{ij=1}^n G_{ij} X_{ji}
\nonumber
\\
&=
\frac{1}{2} \sum_{ijrs=1}^n R^{-1}_{ij,rs} X_{rs} X_{ji}
\nonumber
\\
&=
\frac{1}{2} \bra{X} R^{-1} \ket{X}
\, ,
\label{gBXXbk}
\end{align}
where $\bra{X}=\ket{X}^\dagger$ and $X^\dagger=X$ was used. The main obstruction in using~(\ref{gBXXbk}) is the inversion of $R$ --- even for the very modest case of a bipartite spin-1/2 system, $R$ is of dimension 16, and its inverse is, in general, difficult to compute. The only case we have managed to invert $R$ by brute force is when $t_{ij}=0$ in~(\ref{rhomixedpara}) --- the corresponding metric is
\begin{equation}
\label{gBnpolar}
g_B= \text{diag} \left( \frac{2}{1-16r^2},\frac{2r^2}{1-4r^2},\frac{2r^2 \sin^2\theta}{1-4r^2} \right)
\, ,
\end{equation}
in the coordinate $(r,\theta,\phi)$-basis in which $\mathbf{n}=r(\sin\theta \cos\phi,\sin\theta \sin\phi,\cos\theta)$. Since rotations transform the components of $\mathbf{n}$ and the $t_{ij}$'s separately, we may use~(\ref{gBnpolar}) to compute the total variance of $\rho$. Under a rotation, $\rho$ transforms according to $\rho \rightarrow (U \otimes U) \rho (U^\dagger \otimes U^\dagger)$, with $U \in SU(2)$, so that the fundamental vector field $\hat{S}_A$, corresponding to $S_A \in \mathfrak{su}(2)$, is $\hat{S}_A=-i[\Sigma_A,\rho]$. The total variance of the mixed state~(\ref{rhomixedpara}), with $t_{ij}=0$, is found to be
\begin{align}
\mathbb{D}_{\mathbf{n},\, t_{ij}=0}(\rho)
&=
\sum_{A=1}^3 |\hat{S}_A|_{\text{B}}^2
\nonumber
\\
&=
\frac{4r^2}{1-4r^2}
\, .
\label{totvarmixedn}
\end{align}
On the other hand, the reduced state is 
\begin{equation}
\label{redstate1}
\rho_1=\text{Tr}_2 \rho=\frac{1}{2} I + \mathbf{n} \cdot \boldsymbol{\sigma}
\, ,
\end{equation}
with total variance $\mathbb{D}(\rho_1)=2r^2$, so that the speed excess comes out equal to 
\begin{equation}
\label{semixedn}
\mathbb{F}_{\mathbf{n}, \, t_{ij}=0}(\rho)=\mathbb{D}_{\mathbf{n}, \, t_{ij}=0}(\rho) -2 \mathbb{D}(\rho_1)=
\frac{16 r^4}{1-4r^2}
\, .
\end{equation}
We may, alternatively, take $\mathbf{n}=0$ in~(\ref{rhomixedpara}), keeping all six $t_{ij}$'s as parameters. The inverse of the resulting $R$ takes too long to compute by brute force in Mathematica, so we have to resort to more elaborate methods that can be found in the literature~\cite{Smi:66,Hub:92,Dit:99}. Following the notation in~\cite{Dit:99}, and specifying it to the case at hand, we define the characteristic polynomial of $\rho$,
\begin{equation}
\label{charpolyrho}
\chi(\lambda)=\det(\lambda I-\rho) \equiv \lambda^4+k_1 \lambda^3+k_2\lambda^2+k_3 \lambda +k_4
\, ,
\end{equation}
and the matrices 
\begin{equation}
\label{KNdef}
K =
\left(
\begin{array}{cccc}
0 & 1 & 0 & 0 
\\
0 & 0 & 1 & 0
\\
0 & 0 & 0 & 1
\\
-k_4 & -k_3 & -k_2 & -k_1
\end{array}
\right)
\, ,
\quad
N=
\left(
\begin{array}{cccc}
k_3 & k_2 & k_1 & 1
\\
-k_2 & -k_1 & -1 & 0
\\
k_1 & 1 & 0 & 0
\\
-1 & 0 & 0 & 0
\end{array}
\right)
\, ,
\end{equation}
in terms of which the following matrix $A$ is defined
\begin{equation}
\label{Adef}
A=-\chi(-K^T)^{-1} N
\, .
\end{equation}
Proposition 2 in~\cite{Dit:99} implies, in our notation, that
\begin{equation}
\label{Riprop}
R^{-1}= \sum_{i,j=1}^4 A_{ij} \rho^{i-1} \otimes (\rho^T)^{j-1}
\, ,
\end{equation}
giving for the Bures metric
\begin{align}
g_{\text{B}}(X,X)
&=
\frac{1}{2} \sum_{i,j=1}^4 A_{ij} \bra{X} \rho^{i-1} \otimes (\rho^T)^{j-1} \ket{X}
\nonumber
\\
&=
\frac{1}{2} \sum_{i,j=1}^4 A_{ij} \text{Tr} \left( X \rho^{i-1} X \rho^{j-1} \right)
\, .
\label{gBuresA}
\end{align}
Using this formula we  find that
\begin{equation}
\label{Totalvtij}
\mathbb{F}_{\mathbf{n}=0, \, t_{ij}}(\rho)=\mathbb{D}_{\mathbf{n}=0, \, t_{ij}}(\rho)=
\frac{-256 k_2^2+96 t k_3 -2t^4-48t^2 k_2 -6t^3+288 k_3-64 t k_2 +14 t^2 +240 k_2+30 t-36}{64k_3+64k_2+4t^2+8t-12}
\, ,
\end{equation}
where the $k_i$ are defined in~(\ref{charpolyrho}), $t\equiv t_{11}+t_{22}+t_{33}$, and the first equality is due to the fact that the corresponding reduced state $\rho_1$ is the maximally mixed one, so that its total variance vanishes. We have also been able to compute the total variance for the state $\rho$ in~(\ref{rhomixedpara}) with both $\mathbf{n}$ and $t_{ij}$ nonvanishing, but the corresponding expressions are too long to quote here.
\section{The total acceleration of a curve in $\mathbb{P}$}
\label{OtaociP}
Given a  curve $\rho_t$ in $\mathbb{P}$, parametrized by time, its velocity is (dropping the subscript $t$) $v=\dot{\rho}$, while its acceleration is $a = \rho \ddot{\rho} \tilde{\rho} + \tilde{\rho} \ddot{\rho} \rho$ (see, \eg,~\cite{Chr.etal:23}). In the case where the time evolution of $\rho$ is generated by a hamiltonian $H$ via Sch\"odinger's equation, $\dot{\rho}=-i[H,\rho]$, it can be shown that~\cite{Chr.etal:23}
\begin{equation}
\label{v2a2}
|v|^2=h_2-h_1^2
\, ,
\qquad
|a|^2=h_4-4h_3h_1-h_2^2+8h_2 h_1^2-4h_1^4
\, ,
\end{equation}
where $h_m=\Tr(\rho H^m)$, and the FS metric is being used, so that, \eg, $|v|^2=\Tr(\dot{\rho}^2)/2$, \etc

The quantity $|a|^2$ is a function  on $\mathbb{P}$ that depends on the hamiltonian $H \in \mathfrak{u}$ chosen for the time evolution of the system, $|a|^2=|a(\rho,H)|^2$. Motivated by the analysis in section~\ref{Tvaam}, we specify now the hamiltonian to be an element of $\mathfrak{su}(2) \subset \mathfrak{u}(n+1)$, $H_{\mathbf{n}}=\mathbf{n} \cdot \mathbf{S}^{(s)}$, corresponding to rotating the state around the axis $\mathbf{n}$, and average the squared modulus of the resulting acceleration over all rotation axes, to get the \emph{total (rotational) acceleration} of $\rho$,
\begin{equation}
\label{ttdef}
\langle |a|^2\rangle_{S^2} \equiv \langle |a|^2 \rangle= \int_{S^2} d\mathbf{n} \,  |a(\rho, H_{\mathbf{n}})|^2
\, .
\end{equation}
\subsection{Averaging by integration}
\label{Abi}
 Writing $H=H^A T_A$, with $T_A$ denoting an orthonormal basis in $\mathfrak{su}(2)$, we get from the second of~(\ref{v2a2}),
\begin{align}
\langle |a|^2 \rangle
&=
 \left( < \! T_A T_B T_C T_D \! > -4 < \!  T_A T_B T_C \! >< \! T_D \! > -<\! T_A T_B \! >< \! T_C T_D \! >
\right.
\nonumber
\\
& \qquad
\left. + 8 < \! T_A T_B \! >< \! T_C \! >< \! T_D \! > -4 < \! T_A \! >< \! T_B \! ><\! T_C \! >< \! T_D \! > \right)\int_{S^2} H^A H^B H^C H^D
\, .
\label{vdavexp}
\end{align}
The integrals of monomials in cartesian coordinates over $S^{n-1}$ are given by
\begin{equation}
\label{intccSn}
\int_{S^{n-1}} x_1^{m_1} \ldots x_n^{m_n}=\frac{(n-2)!! \prod_{i=1}^n (m_i-1)!!}{n-2+\sum_{i=1}^n m_i}
\, ,
\end{equation}
where $\int_{S^{n-1}} 1=1$ and the above $m_i$ are all even (if any $m_i$ is odd, the integral is zero). In our case, we only need 
\begin{equation}
\label{twont}
\int_{S^2} x_i^2 x_j^2=\frac{1}{15} \quad \text{for } i \neq j
\, ,
\qquad
\int_{S^2} x_i^4=\frac{1}{5}
\, .
\end{equation}
For spin 1, one finds
\begin{equation}
\label{s1vd2av}
\langle|a|^2 \rangle
=
\frac{1459+1344 \cos(2\alpha)+140 \cos(4\alpha) +128 \cos(6\alpha) +\cos(8\alpha)}{
60 \left(3+\cos(2\alpha) \right)^4}
\, ,
\end{equation}
where $2\alpha$ is the angle between the two Majorana stars --- a plot appears in Fig.~\ref{vd2s1Plot_Fig} (left).
\setlength{\figurewidth}{\textwidth}
\begin{figure}
\centerline{%
\hfill
\raisebox{0\totalheight}{%
\includegraphics[angle=0,width=.40\figurewidth]%
{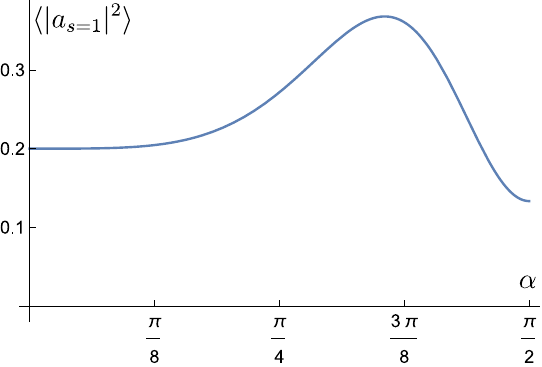}%
}
\hfill
\raisebox{.0\totalheight}{%
\includegraphics[angle=0,width=.40\figurewidth]%
{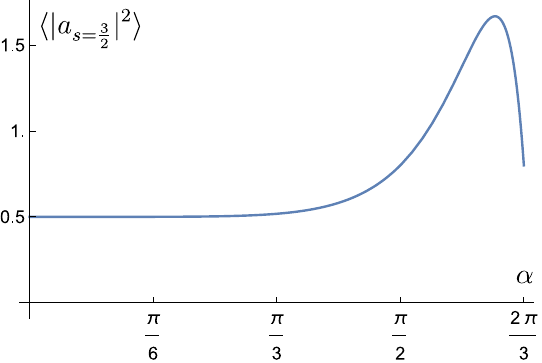}%
}
\hfill
}
\caption{
Left: Plot of $\langle |a|^2 \rangle$ \emph{vs.} $\alpha$ ($s=1$, $\alpha$ is the half-angle between the two Majorana stars).
Right: Plot of $\langle  |a|^2  \rangle$ \emph{vs.} $\alpha=\beta=\gamma$ ($s=3/2$, $\alpha$ is the angle between any two of the three Majorana stars).
}
\label{vd2s1Plot_Fig}
\end{figure}
For spin 3/2 things get considerably more complicated. We have computed $\langle |a|^2\rangle_{S^2}$ in terms of the angles $\alpha$, $\beta$, $\gamma$ between the three Majorana stars. The expression simplifies  along the diagonal ($\alpha=\beta=\gamma$),
\begin{align}
\langle |a|^2 \rangle_{\alpha=\beta=\gamma}
&=
\frac{5774 \cos (\alpha )+1793
   \cos (2 \alpha )+1027 \cos
   (3 \alpha )+82 \cos (4
   \alpha )-17 \cos (5 \alpha
   )-\cos (6 \alpha
   )+2862}{1440 (\cos (\alpha
   )+1)^4}
\, .
\label{s32vd2av}
\end{align}
A plot appears in Fig.~\ref{vd2s1Plot_Fig} (right), while a contour plot of the full function is shown in Fig.~\ref{vd2s32Plot_Fig}.
\setlength{\figurewidth}{\textwidth}
\begin{figure}
\centerline{%
\raisebox{0\totalheight}{%
\includegraphics[angle=0,width=.40\figurewidth]%
{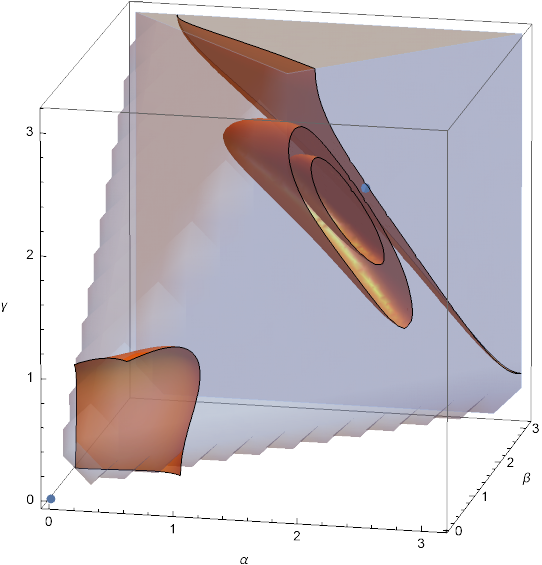}%
}
}
\caption{
ContourPlot of $\langle  |a|^2  \rangle$ \emph{vs.} $\alpha$, $\beta$, $\gamma$ ($s=3/2$, the constellation shape is parametrized in terms of the angles between the stars). Shown are the level surfaces at $.51$ (lower left and upper right surfaces), $1.3$ (outer closed surface), $1.5$ (inner closed surface). The blue dots correspond to the coherent state $\alpha=\beta=\gamma=0$ (lower left corner) and the GHZ state $\alpha=\beta=\gamma=2\pi/3$ (between the two uppermost surfaces).
}
\label{vd2s32Plot_Fig}
\end{figure}

A rather long calculation, the details of which are given in the appendix, results in the following expression for the average norm squared of the (rotational) acceleration for a general spin $s$,
\begin{align}
\langle |a|^2 \rangle 
= & \lambda_1 + \lambda_2 |\bm{\rho}_1|^2 +\lambda_3 |\bm{\rho}_2|^2 + \lambda_4 c_{1 N_1 1 N_2}^{2 N_1+ N_2} \rho_{2 N_1+N_2}
\rho_{1 N_1}^* \rho_{1 N_2}^* 
+ \lambda_5 |\bm{\rho}_1|^4 
\, ,
\label{Fin.ans}
\end{align}
where $\bm{\rho}_L = (\rho_{LL} , \dots , \rho_{L-L})$, with $\rho_{LM}=\langle T^{\dagger}_{LM} \rangle$, and
\begin{align}
\lambda_1 
&= \frac{s(s+1)(2s-1)(2s+3)}{45} 
\, ,
\label{lambda1def}
\\
\lambda_2 
&= 
\left\{
\begin{array}{c c}
\frac{4}{27}s^2(s+1)^2(2s+1) & s \leq 1/2 
\\
\frac{4}{135}s(s+1)(2s+1)(s^2+s+3) & s > 1/2
\end{array}
\right. 
\, ,
\label{lambda2def}
\\
\lambda_3 
&=
 -\frac{1}{225}s(s+1)(2s-1)(2s+1)(2s+3) 
 \, ,
 \label{lambda3def}
 \\
\lambda_4 
&=
 \frac{8}{45}\sqrt{\frac{2}{15}}s(s+1)(2s+1)\sqrt{s(s+1)(2s-1)(2s+1)(2s+3)} 
 \, ,
 \label{lambda4def}
\\
\lambda_5 
&=
 -\frac{4}{45}s^2 (s+1)^2(2s+1)^2 
 \, .
 \label{lambda5def}
\end{align}
We test our Eq. \eqref{Fin.ans} with the spin-1 state $\ket{\psi}= (\cos^2\frac{\alpha}{2},0,-\sin^2\frac{\alpha}{2})/\sqrt{\cos^4\frac{\alpha}{2}+\sin^4\frac{\alpha}{2}}$, and recover the expression in~(\ref{s1vd2av}). Putting $\cos A=\cos \frac{\alpha}{2}/\sqrt{\cos^4\frac{\alpha}{2}+\sin^4\frac{\alpha}{2}}$, so that $\ket{\psi}=(\cos A, 0 , -\sin A)$, we find
\begin{equation}
\langle |a|^2  \rangle = 
\frac{1}{30} \left( 8+ \cos(4A)-3 \cos(8A) \right)
 \, .
\label{Case.s1}
\end{equation} 
Table \ref{lambdai} contains the values of $\lambda_i$, $i=1,\ldots,5$, for $s=1/2,\dots, 2$.
\begin{table}
\begin{tabular}{c | c c c c c }
$s$ & $\lambda_1$ & $\lambda_2$ & $\lambda_3$ & $\lambda_4$ & $\lambda_5$ 
\\
\hline
$\rule{0ex}{3ex} \frac{1}{2}$ & 0 & $\frac{1}{6}$ & 0 & 0 & $-\frac{1}{5}$
\\
$\rule{0ex}{3ex} 1$ & $\frac{2}{9}$ & $\frac{8}{9}$ & $-\frac{2}{15}$ & $\frac{32}{15}$ & $-\frac{16}{5}$
\\
$\rule{0ex}{3ex} \frac{3}{2}$ & 1 & 3 & $-\frac{4}{5}$ & $16\sqrt{\frac{2}{3}}$ & $-20$
\\
$\rule{0ex}{3ex} 2$ & $\frac{14}{5}$ & 8 & $-\frac{14}{5}$ & $32 \sqrt{\frac{7}{3}}$ & $-80$
\end{tabular}
\caption{\label{lambdai}List of the $\lambda_i$ values in Eqs.~(\ref{lambda1def}) -- (\ref{lambda5def}) for $i=1,\ldots,5$, and $s=1/2,1,3/2,2$.}
\end{table}
Note that (\ref{Fin.ans}) implies that all 2-anticoherent spin states, for which $\boldsymbol{\rho}_1=\boldsymbol{\rho}_2=0$,  have the same total acceleration $\langle |a|^2  \rangle = \lambda_1$.
Also, for a spin-$s$  coherent state $\ket{n}$ we find
\begin{equation}
\label{ttSC}
\langle |a|^2 \rangle_{\ket{n}}=\frac{1}{45} s\left( 8s^2(s+1)-4s-3 \right)
\, ,
\end{equation}
which is seen to grow asymptotically, for large $s$,  like $s^4$.
\subsection{Averaging using spherical designs}
\label{Aasd}
While for a general function $f \colon S^2 \rightarrow \mathbb{R}$ integration is necessary in order to compute its average, the polynomial functions that show up in, \eg, (\ref{vdavexp}), can in fact be averaged by sampling them over an appropriate  finite set of points. Our first relevant concept is that of a  \emph{spherical $t$-design} in dimension $d$, defined as a set of points $\{p_i \}$, $i=1,\ldots,N$, on $S^d$ such that the average of any polynomial $f$ of degree $t$  or less (in the cartesian coordinates $\{x^1,\ldots,x^{d+1} \}$ of the ambient $\mathbb{R}^{d+1}$) over the set coincides with the average over $S^d$, \ie,
\begin{equation}
\label{Sddef}
\frac{1}{N}\sum_{i=1}^N f(p_i) =\frac{1}{|S^d|} \int_{S^d} f \td \Omega
\, ,
\end{equation}
where $|S^d|=\int_{S^d} \td \Omega$, and $\td \Omega$ is the euclidean measure on $S^d$~\cite{Del.Goe.Sei:77,Sey.Zas:84}. There are two features of the integrand in, \eg, (\ref{vdavexp}), that call for a refinement of the above concept: the polynomial in question is homogeneous, and of an even degree, as fit for a squared modulus. The appropriate concept then is that of a \emph{spherical $(t,t)$-design} in dimension $d$, which is the specification of the previous definition to the case of a homogeneous polynomial of degree $2t$~\cite{Hug.Wal:21}. An obvious property of spherical designs (of either of the above types) is that it  can be rotated arbitrarily, remaining a spherical design. We have already seen a spherical design entering the discussion above: the average of the modulus squared of the rotational velocity,  which is homogeneous quadratic in the components of the hamiltonian $H$,  came out as the average of its value for rotations around the three coordinate axes. In the parlance introduced above, this follows from
 the statement that any three mutually orthogonal directions on $S^2$ furnish a spherical $(1,1)$-design.  The squared modulus of the acceleration in~(\ref{vdavexp}) is homogeneous quartic in the components of the hamiltonian, so the integral over $S^2$ in that expression may also be computed by averaging over a spherical $(2,2)$-design. A minimal such design (\ie, with the minimum possible number of points) is given in~\cite{Hug.Wal:21} --- it consists of the six equiangular lines which go through the vertices of an icosahedron (for each such line, any of the two half-lines emanating from the origin may be chosen, since the functions being averaged are symmetric on antipodal points of $S^2$). Choosing the vertices with positive $z$-coordinate, among the icosahedron vertices provided by Mathematica 13.2, we find the six-point spherical $(2,2)$-design 
 \begin{equation}
 \label{s22d}
 p_1=(0,0,1)
 \, ,
 \quad
 p_2=
 (\frac{2}{\sqrt{5}},0,\sigma)
 \, , 
 \quad
p_3= (-\mu^2,-\nu,\sigma)
  \, , 
 \quad
p_4=
 (-\mu^2,\nu,\sigma)
  \, , 
 \quad
 p_5=
 (\nu^2,-\mu,\sigma)
  \, , 
 \quad
 p_6=
 (\nu^2,\mu,\sigma)
  \, , 
 \end{equation}
where $\mu \equiv \sqrt{(5+\sqrt{5})/10}$,  $\nu \equiv \sqrt{(5-\sqrt{5})/10}$, $\sigma=1/\sqrt{5}$, \ie, 
 with one point  at the north pole, and the other five on the vertices of a regular pentagon at $z=1/\sqrt{5}$. For any value of spin $s$, the hamiltonians $H_i=p_i \cdot \mathbf{S}^{(s)}$ are to be used in the second of~(\ref{v2a2}) to compute the corresponding squared moduli $a_i^2$, the average of which gives $\langle a^2 \rangle$ --- we have checked~(\ref{s1vd2av}) doing just that for $s=1$.
\section{Acceleration Addition Law}
\label{AAL}
\subsection{Acceleration addition law for pure bipartite separable states}
\label{SectionAcceleration}
We look here for the addition law for the norm squared of the acceleration of a separable state  $\rho=\rho_1\otimes\rho_2$ --- the corresponding result for the norm squared of the velocity is given in (\ref{spadlawPytha}). With the acceleration given by~(\ref{accdef}),
where
\begin{align}
	\label{Ec:ddotrho}
	\dot{\rho}
	&=
	\dot{\rho}_1 \otimes \rho_2+\rho_1\otimes\dot{\rho}_2,\nonumber
	\\
	\ddot{\rho}
	&=
	\ddot{\rho}_1\otimes \rho_2 + 2 \dot{\rho}_1\otimes\dot{\rho}_2+\rho_1\otimes \ddot{\rho}_2
	\, ,
\end{align}
we get
\begin{equation}
	\label{Eq:accCuad}
	|a|^2
	=
	\frac{1}{2}\text{Tr}(a^2) =\ldots =
	\text{Tr}(\rho\ddot{\rho}^2-\rho\ddot{\rho}\rho\ddot{\rho}),
\end{equation}
where we used the cyclicity of the trace and the fact that $\rho\tilde{\rho}=\tilde{\rho}\rho=0$. Some further gymnastics reveal that
\begin{align}
	\label{Eq:firstTerm}
	\text{Tr}(\rho\ddot{\rho}^2)
	&=
	\text{Tr}\left(\rho_1\ddot{\rho}_1^2
	+
	\rho_2\ddot{\rho}_2^2\right)
	+
	2\text{Tr}\left(\rho_1\ddot{\rho}_1\right)\text{Tr}\left(\rho_2\ddot{\rho}_2^2\right)
	+
	4\text{Tr}\left(\rho_1\dot{\rho}_1^2\right)\text{Tr}\left(\rho_2\dot{\rho}_2^2\right)
	\, ,
	\\
	\text{Tr}(\rho\ddot{\rho}\rho\ddot{\rho})
	&=
	\text{Tr}\left(
	\rho_1\ddot{\rho}_1\rho_1\ddot{\rho}_1+\rho_2\ddot{\rho}_2\rho_2\ddot{\rho}_2
	\right)
	+
	2\text{Tr}\left(
	\rho_1\ddot{\rho}_1
	\right)\text{Tr}\left(
	\rho_2\ddot{\rho}_2
	\right)
	+
	4 \text{Tr}\left(
	\rho_1\dot{\rho}_1\rho_1\dot{\rho}_1
	\right)\text{Tr}\left(
	\rho_2\dot{\rho}_2\rho_2\dot{\rho}_2
	\right)
	\, .
\end{align}
Noting that
\begin{equation}
	\text{Tr}(\rho_1\dot{\rho}_1\rho_1\dot{\rho}_1)
	=
	\text{Tr}(\rho_1\dot{\rho}_1)^2=0
	\, ,
	\qquad
	\text{Tr}(\rho\dot{\rho}^2)=\frac{1}{2}\text{Tr}(\dot{\rho}^2)
	\, ,
\end{equation}
 we finally get
\begin{equation}
	\label{Eq:AdditionLawAcc}
	|a|^2=|a_1|^2+|a_2|^2+4 |v_1|^2 |v_2|^2
	\, ,
\end{equation}
showing that the speed of the subparts contributes to the acceleration of the whole --- in particular, a bipartite system can have acceleration even though its subparts do not. For example, consider a two spin-1/2  system in the coherent state in the $x$-direction, 
\begin{equation}
	\ket{\psi}=\ket{+,\hat{x}}\otimes\ket{+,\hat{x}},
\end{equation}
where $\ket{+,\hat{x}}=\frac{1}{\sqrt{2}}(1,1)^T$ in the eigenbasis of $S_z$. If $\ket{\psi}$ is evolved by rotating about the  $z$ axis, one easily finds that $|a_1|^2=|a_2|^2=0$, since the subsystem stars move along great circles on the Bloch sphere, while
\begin{equation}
	\ket{\psi(t)}=\left(
	\begin{matrix}
		\frac{e^{-i t}}{2} &
		\frac{1}{2} &
		\frac{1}{2} &
		\frac{e^{i t}}{2}
	\end{matrix}
	\right)^T
	\quad
	\Rightarrow
	\quad
	a=[\rho,[\rho,\ddot{\rho}]]=\frac{1}{4}\left(
	\begin{array}{cccc}
		-1 & 0 & 0 & -e^{-2 i t} \\
		0 & 1 & 1 & 0 \\
		0 & 1 & 1 & 0 \\
		-e^{2 i t} & 0 & 0 & -1 \\
	\end{array}
	\right),
\end{equation}
so that
$|a|^2=\frac{1}{2}\text{Tr}(a^2)=\frac{1}{4}$, \ie, the entire acceleration of the composite system is due to the speed of its subsystems, each of which moves along geodesics of the FS metric. 

Another interesting consequence of~(\ref{Eq:AdditionLawAcc}) is that the only separable state curves that are geodesics (characterized by $|a|=0$) are those for which one of the parts does not evolve at all, while the other follows a geodesic.
\subsection{Acceleration of mixed bipartite states}
\subsubsection{Parallel transport in the space of density matrices}%
\label{subsec:parallelTransport}
Consider two vector fields $X$, $Y$, defined in a neighborhood of a density matrix $\rho$ and denote by $\overline X$,
$\overline Y$ their horizontal lifts in $S$. Then, as we prove for completeness below,
\begin{equation}
  \begin{split}
 \nabla_{X} Y = \pi_{*}(\overline \nabla_{\overline X} \overline Y),
  \end{split}
  \label{eq:ptFormula}
\end{equation}
where $\nabla$ denotes the Levi-Civita connection of the space of density matrices, and $\overline \nabla$ the one of $S$.
Geometrically, (\ref{eq:ptFormula}) states  that the parallel transport of the vector $Y$ along $X$ can be obtained by effecting
the corresponding parallel transport in $S$ using horizontal lifts, and then projecting back to the space of density matrices.

By definition, proving that (\ref{eq:ptFormula}) gives the Levi-Civita connection reduces to showing that for $\nabla$ is compatible
with the metric and that it is torsion free.

To verify compatibility with the metric, it is enough to prove that, for vector fields $X,Y,Z$,
the equality $g(Y+  \epsilon \nabla_X Y, Z+  \epsilon \nabla_X Z)= g(Y, Z)+ \mathcal{O}(\epsilon^{2})$ holds. For (\ref{eq:ptFormula}), this requires proving that
\begin{equation}
  \begin{split}
g(Y,Z)=g(\pi_{*}[\overline Y+ \epsilon \overline \nabla_{\overline X} \overline Y], \pi_{*}[\overline Z+ \epsilon \overline \nabla_{\overline X} \overline Z])
    +\mathcal{O}(\epsilon^{2})
    \, .
  \end{split}
\end{equation}
Recall that $g$  can be obtained from $\overline g$ by considering the horizontal components of the involved vectors. Therefore,
\begin{equation}
  \begin{split}
    \overline g(\overline Y+ \epsilon \overline \nabla_{\overline X} \overline Y, \overline Z+\epsilon  \overline \nabla_{\overline X} \overline Z)&=
    g(\pi_{*}[\overline Y+\epsilon  \overline \nabla_{\overline X} \overline Y], \pi_{*}[\overline Z+\epsilon  \overline \nabla_{\overline X} \overline Z])+
    \overline g(\overline Y^{V}+\epsilon  (\overline \nabla_{\overline X} \overline Y)^{V}, \overline Z^{V}+\epsilon  (\overline \nabla_{\overline X} \overline Z)^{V})
    \\&=
    g(\pi_{*}[\overline Y+\epsilon  \overline \nabla_{\overline X} \overline Y], \pi_{*}[\overline Z+\epsilon  \overline \nabla_{\overline X} \overline Z])+
    \epsilon^{2}\overline g( [\overline \nabla_{\overline X} \overline Y]^{V}, [\overline \nabla_{\overline X} \overline Z]^{V})
  \end{split}
  \label{eq:gupvsgdown}
\end{equation}
where ${W}^{V}$ denotes the vertical component of $W$, and we used the fact that  $\overline Y^{V}=\overline Z^{V}=0$
holds by definition to obtain the last equality.
Finally, by writing $g(Y,Z)= \overline g(\overline Y,\overline Z)$ and noting that  $\overline \nabla$ is compatible with $\overline g$
we conclude,
\begin{equation}
  \begin{split}
    g(Y,Z)= \overline g(\overline Y,\overline Z)=\overline g(\overline Y+ \epsilon \overline \nabla_{\overline X} \overline Y, \overline Z+\epsilon  \overline \nabla_{\overline X} \overline Z)&=
    g(\pi_{*}[\overline Y+\epsilon  \overline \nabla_{\overline X} \overline Y], \pi_{*}[\overline Z+\epsilon  \overline \nabla_{\overline X} \overline Z])
                                                                                                                                                                          + \mathcal O(\epsilon^{2}),
  \end{split}
\end{equation}
where we used (\ref{eq:gupvsgdown}) for the last line. This proves that (\ref{eq:ptFormula}) is compatible with $g$.

The second requirement, is that (\ref{eq:ptFormula}) is torsion free. This is equivalent to the equality
$\pi_{*}(\overline \nabla_{\overline X} \overline Y-\overline \nabla_{\overline y} \overline X)=[X,Y]$ 
for all vector fields $X,Y$. As  is well-known from the theory of fiber bundles, the projection $\pi_{*}$ is compatible with
the commutator of horizontal fields, so $\pi_{*}[\overline X, \overline Y] = [\pi_{*}\overline X,\pi_{*} \overline Y]$ holds.
Therefore we have,
\begin{equation}
  \begin{split}
[X,Y] =[\pi_{*}\overline X,\pi_{*} \overline Y]= \pi_{*}[\overline X, \overline Y]=
    \pi_{*}(\overline \nabla_{\overline X} \overline Y -\overline \nabla_{\overline y} \overline X ),
  \end{split}
\end{equation}
where we used that $\overline \nabla$ is torsion free, proving  that (\ref{eq:ptFormula}) does indeed gives the Levi-Civita connection $\nabla$.

As an immediate application of the previous result, we can compute the intrinsic acceleration
of a curve of density matrices $\rho(t)$ with tangent vector $X(t)$,
\begin{equation}
  \begin{split}
a =  \nabla_{X} X = \pi_{*}  \overline \nabla_{\overline X}\overline  X = \pi_{*}\overline a,
  \end{split}
  \label{eq:accUpVsDown}
\end{equation}
meaning that we can compute the corresponding acceleration in $S$ and then simply project it with $\pi_{*}$ to obtain $a$.
For the Bures metric, this procedure becomes particularly simple, since $S$ is embedded in $\mathcal{HS}^{*}$, and the latter is Euclidean --- we work out the details in what follows.

Consider an horizontal curve $A(t)$ in $S$. Horizontality implies that we can write $\dot A= G A$, with $G$ hermitian.
Its intrinsic acceleration is simply the orthogonal projection of $\ddot A$ to the
sphere $S$, as shown in
(\ref{addproj}).
This result, together with  (\ref{eq:accUpVsDown}),
implies that the intrinsic acceleration of $\rho=A A^{\dagger}$ is given by
\begin{equation}
  \begin{split}
a = \pi_{*} (\ddot A-(1/2)\Tr(\ddot A A^{\dagger} + A \ddot A^{\dagger})A)= \ddot A A^{\dagger} + A \ddot A^{\dagger}- \Tr(\ddot A A^{\dagger} + A \ddot A^{\dagger}) \rho.
  \end{split}
\end{equation}
We proceed to write everything in terms of $\rho$ and $G$. Since $\dot A= G A$,
\begin{equation}
  \begin{split}
\ddot A = \dot G A + G \dot A= \dot G A + G^{2}  A,
  \end{split}
\end{equation}
so,
\begin{equation}
  \begin{split}
a &= (\dot G A + G^{2}  A) A^{\dagger} + A (A^{\dagger} \dot G  + A^{\dagger}G^{2}  )- \Tr(\ddot A A^{\dagger} + A \ddot A^{\dagger}) \rho
    \\&
= (\dot G  + G^{2}  ) \rho + \rho ( \dot G  + G^{2}  )- \Tr(\ddot A A^{\dagger} + A \ddot A^{\dagger}) \rho
  \end{split}
\end{equation}
On the other hand, by
(\ref{dPGP}),
$\dot \rho = G \rho+ \rho G$, so
\begin{equation}
  \begin{split}
\ddot \rho &= \dot G \rho+ \rho \dot G+ G \dot \rho+ \dot \rho G
          \\
 &= \dot G \rho+ \rho \dot G+ G (G \rho + \rho G)+ (G \rho + \rho G) G
          \\&=
 \dot G \rho+ \rho \dot G+  G^{2} \rho + \rho G^{2}+2G \rho G
          \\&=
 a+2G \rho G+\Tr(\ddot A A^{\dagger} + A \ddot A^{\dagger}) \rho\,.
  \end{split}
\end{equation}
Therefore,
\begin{equation}
  \begin{split}
    a=
\ddot \rho-2G \rho G- \Tr(\ddot A A^{\dagger} + A \ddot A^{\dagger}) \rho\,.
  \end{split}
\end{equation}
By noting that $\Tr \, a=\Tr\, \ddot \rho=0$, and taking the trace in the previous expression,
we obtain the equality $\Tr(\ddot A A^{\dagger} + A \ddot A^{\dagger})=-2\Tr(G \rho G)$,
so,
\begin{equation}
  \begin{split}
a= \ddot \rho-2G \rho G+ 2\Tr(G \rho G) \rho.
  \end{split}
  \label{eq:accelerationBures}
\end{equation}
Equation (\ref{eq:accelerationBures}) is the main result of this section.
For the particular case of  Hamiltonian evolution,
\begin{equation}
  \begin{split}
\ddot \rho = -[H,[H, \rho]],
  \end{split}
\end{equation}
so we have,
\begin{equation}
  \begin{split}
    a= -[H,[H, \rho]] -2G \rho G+ 2 \Tr(G \rho G) \rho,
  \end{split}
\end{equation}
where $G$ is uniquely determined by,
\begin{equation}
  \begin{split}
-i [H, \rho] = G \rho+ \rho G.
  \end{split}
\end{equation}
provided all eigenvalues of $\rho$ are positive
\cite{Chr.etal:23}.
\subsubsection{Working with the projection $\tilde \pi (A)= A A^{\dagger}/\mathrm{Tr}(A A^{\dagger})$}
When working with a curve $A(t)$ in $\mathcal{HS}^{*}$ with the idea of projecting it to a curve $\rho(t)$,
it is often useful to consider the projection $\tilde \pi (A)= A A^{\dagger}/\Tr(A A^{\dagger})$ instead of $\pi$, so that
we are not required to work with matrices $A$ such that $\Tr(A A^{\dagger})=1$. Below we provide the details
to work with $\tilde \pi$.

Consider $A(t)$ a $\pi$-horizontal curve (so, by (\ref{dPGP}), $\dot A= \tilde G A$ with $\tilde G$ hermitian) and considered the projected curve $\rho=\tilde \pi (A)$.
Note that,
\begin{equation}
  \begin{split}
    \dot \rho &= \frac{d}{dt}\left(\frac{AA ^{\dagger}}{\Tr(A A^{\dagger})}\right)=
    (\tilde G \rho + \rho \tilde G)-2\frac{\Tr(\tilde G \rho)}{\Tr(A A^{\dagger})} A A^{\dagger}
    \\&
    =(\tilde G \rho + \rho \tilde G)-2{\Tr(\tilde G \rho)} \rho
    =\left(\tilde G- \Tr(\tilde G \rho) I\right) \rho + \rho\left(\tilde G- \Tr(\tilde G \rho) I\right)
  \end{split}
\end{equation}
where $I$ denotes the identity matrix and we used
the expressions $(d/dt)(A A^{\dagger})= \Tr(A A^{\dagger}) (\tilde G \rho + \rho \tilde G)$ and $(d/dt)\Tr(A A^{\dagger})= 2\Tr(A A^{\dagger})\Tr(\tilde G \rho)$
for the first equality. If we define $G=\tilde G- \Tr(\tilde G \rho) I$, we recover
(\ref{dPGP}),
and we can compute the Bures metric using (\ref{eq:altgg}).

Note that in this case, neither $G$ satisfies the relation $\dot A= G A$, nor does the size
of $\dot{\rho}$ coincides with the one of $\dot A$ (not even after normalizing $\dot A$  by dividing it by $[\Tr(A A^\dagger)]^{1/2}$),
\begin{equation}
  \begin{split}
\frac{\overline g(\dot A,\dot A)}{\Tr(A A^{\dagger})}= \frac{\Tr(\dot A \dot A^{\dagger})}{\Tr(A A^{\dagger})} =
\Tr(\tilde G \rho \tilde G) = \Tr (\rho \tilde G^{2} )\,,
  \end{split}
\end{equation}
where we used $\dot A= \tilde G A$ to obtain the second equality.
By writing $\tilde G$ in terms of $G$ and using $\Tr(G \rho)=0$ we conclude,
\begin{equation}
  \begin{split}
\frac{\overline g(\dot A,\dot A)}{\Tr(A A^{\dagger})} = g(\dot \rho,\dot \rho)+[\Tr(\rho \tilde G)]^{2} \geq g(\dot \rho,\dot \rho)\,.
  \end{split}
  \label{eq:inequalityUpDwn}
\end{equation}
The reason for the difference between the sizes of $\dot \rho$ and  that of $\dot A/ [\Tr(A A^{\dagger})]^{1/2}$
is that the latter has a vertical component with respect to $\tilde \pi$. Recall that we assumed that  $A(t)$  was an horizontal
vector for $\pi$, not for $\tilde \pi$.
For the projection $\tilde \pi$, the horizontal vectors $V$, besides having to satisfy the condition
$V=\tilde GA$ (with $\tilde G$ hermitian), also have to satisfy the condition $\Tr(VA^{\dagger})=\Tr(\tilde G A A^{\dagger})=0$.
The reason for this  is that curves of the form $\gamma(t)=t A$ are horizontal for $\pi$, but vertical for $\tilde \pi$
--- indeed they get projected by $\tilde \pi$ to a single point. If we assume that  $V$ is orthogonal to this kind of curves, we
conclude that $\tilde \pi$-horizontal vectors $V$ satisfy additionally the condition $\Tr(VA^{\dagger})=0$.

If we work with $\tilde \pi$-horizontal curves, by the definition of $G$, we immediately have $\tilde G=G$ and
the equality in (\ref{eq:inequalityUpDwn}) holds, as expected.

Finally, note that $\tilde \pi$-horizontal curves live in the sphere where the term $\Tr(A A^{\dagger})$ is constant,
\begin{equation}
  \begin{split}
\frac{d}{dt}\Tr (A A^{\dagger})= \Tr (\dot A A^{\dagger}+ A \dot A^{\dagger})=2\Tr (\tilde G A A^{\dagger})=0\,.
  \end{split}
\end{equation}
In particular,  if $\Tr(A A^{\dagger})=1$ at the initial time, we are automatically in the usual case of the Bures metric.

So, in conclusion, if we work with $\tilde \pi$-horizontal curves, we can compute the Bures metric like we did in the previous section,
since $\tilde G=G$ in this case.
If we do not, we have to compute it using the operator $G$ defined as $G=\tilde G- \Tr(\tilde G \rho) I$, where we recall the reader that $\tilde G$ is given by
the equality $\dot A=\tilde G A$.
\subsubsection{Explicit expression for the Christoffel symbols of the Bures metric}%
\label{sec:numcheck}
As we show below, equation (\ref{eq:accelerationBures}) allows to find an expression for the Christoffel symbols that does not
involve the derivatives of the operator $G$. Since computing $G$ is generally a numerically  demanding task, it is useful to avoid
computing its derivatives, as it would be done in the usual approach to compute the Christoffel symbols.

Suppose we parametrize the space of density matrices using coordinates $\mu$. The expression for
the acceleration of a curve is,
\begin{equation}
  \begin{split}
a = \ddot \mu \rho_{\mu}+ \Gamma^{\alpha}{}_{\mu \nu} \dot \mu \dot \nu \rho_{\alpha},
  \end{split}
  \label{eq:christoffelAcc}
\end{equation}
where we defined $\rho_{\mu}= \partial_{\mu} \rho$. On the other hand,
\begin{equation}
  \begin{split}
\ddot \rho = \frac{d}{dt} (\dot \mu \rho_{\mu})= \ddot \mu \rho_{\mu}+ \dot \mu \dot \nu \rho_{\mu \nu}.
  \end{split}
\end{equation}
where $\rho_{\mu \nu}= \partial_{\mu \nu} \rho$. By comparing (\ref{eq:christoffelAcc})  with (\ref{eq:accelerationBures}),
we conclude the following expression,
\begin{equation}
  \begin{split}
\dot \mu \dot \nu \rho_{\mu \nu} -2G \rho G+2\Tr(G \rho G) \rho = \Gamma^{\alpha}{}_{\mu \nu} \dot \mu \dot \nu \rho_{\alpha}.
  \end{split}
  \label{eq:christtofelvsBures}
\end{equation}
If we denote $G_{\mu}$ the operator defined analogously to $G$ in (\ref{dPGP}),
\begin{equation}
  \begin{split}
\rho_{\mu} = G_{\mu} \rho + \rho G_{\mu}.
  \end{split}
\end{equation}
then,
\begin{equation}
  \begin{split}
2 G \rho G &=2 \dot \mu \dot \nu G_{\mu} \rho G_{\nu}= \dot \mu \dot \nu (G_{\mu} \rho G_{\nu}+G_{\nu} \rho G_{\mu}),
          \\
 2\Tr(G \rho G) &= \dot \mu \dot \nu \Tr(G_{\mu} \rho G_{\nu}+G_{\nu} \rho G_{\mu}),
  \end{split}
\end{equation}
so that (\ref{eq:christtofelvsBures}) becomes,
\begin{equation}
  \begin{split}
\dot \mu \dot \nu (\rho_{\mu \nu} -G_{\mu} \rho G_{\nu}-G_{\nu} \rho G_{\mu}+ \Tr[G_{\mu} \rho G_{\nu}+G_{\nu} \rho G_{\mu}] \rho) =\dot \mu \dot \nu  \Gamma^{\alpha}{}_{\mu \nu} \rho_{\alpha}.
  \end{split}
\end{equation}
Since this expression holds for arbitrary $\dot \mu, \dot \nu$,
\begin{equation}
  \begin{split}
\rho_{\mu \nu} -G_{\mu} \rho G_{\nu}-G_{\nu} \rho G_{\mu}+ \Tr(G_{\mu} \rho G_{\nu}+G_{\nu} \rho G_{\mu})  \rho= \Gamma^{\alpha}{}_{\mu \nu} \rho_{\alpha}.
  \end{split}
\end{equation}
To solve for $\Gamma$, we compute the inner product of both sides of the equation with $\rho_{\beta}$,
\begin{equation}
  \begin{split}
g(\rho_{\mu \nu} -G_{\mu} \rho G_{\nu}-G_{\nu} \rho G_{\mu}+ \Tr[G_{\mu} \rho G_{\nu}+G_{\nu} \rho G_{\mu}] \rho, \rho_{\beta})= \Gamma^{\alpha}{}_{\mu \nu} g(\rho_{\alpha}, \rho_{\beta})=\Gamma_{\beta\mu \nu} .
  \end{split}
\end{equation}
We can compute the l.h.s.\ using a direct approach --- by writing $g$ in terms of traces with the help of
(\ref{grho12}),
\begin{equation}
  \begin{split}
g(\rho_{\mu \nu} -G_{\mu} \rho G_{\nu}-G_{\nu} \rho G_{\mu}+ \Tr[G_{\mu} \rho G_{\nu}+G_{\nu} \rho G_{\mu}] \rho, \rho_{\beta}) &= \frac{1}{2} \Tr ( G_{\beta}(\rho_{\mu \nu} -G_{\mu} \rho G_{\nu}-G_{\nu} \rho G_{\mu}) ),
  \end{split}
\end{equation}
where we used that $\Tr(G_{\beta} \rho)=\Tr(\dot \rho)/2=0$ to get rid of the last term.
By equating these equations we conclude finally,
\begin{equation}
  \begin{split}
\Gamma_{\beta\mu \nu} =
 \frac{1}{2} \Tr ( G_{\beta}[\rho_{\mu \nu} -G_{\mu} \rho G_{\nu}-G_{\nu} \rho G_{\mu}] ).
  \end{split}
\end{equation}
The previous result was verified numerically for a spin $1$ system at several random points.
\section{Kinematical quantities, purity and quantum correlations}
\label{KQPaQC}
In this section we explore numerically the relation between averaged velocities and accelerations (both of the whole system and of the corresponding subsystems), different measures of quantum correlations (concurrence, negativity and geometric quantum discord) and the purity of the states (von Neumann entropy and linear entropy, also known as $1$-anticoherence measure). We restrict our analysis  to the case of bipartite systems of qubits in a symmetric state evolving through rotations. Even though far from comprehensive, this numerical approach reveals interesting relations that we plan  to explore further in future work.

Let us recall some definitions of quantities explored in this section. Concurrence and negativity are two well-known measures of entanglement, while von Neumann entropy and linear entropy quantify how mixed  a quantum state is. Geometric quantum discord captures information about the state when a measurement is performed in one of the subsystems that compose the entire system. Consider an arbitrary two-qubit state,
\begin{equation}
\label{state2q}
    \rho= 
    \frac{1}{4}
    \left(
    I_2 \otimes I_2
    +
    x \cdot \sigma \otimes I_2
    +
    I_2 \otimes y \cdot \sigma
    +
\sum_{i,j=1}^{3}T_{ij} \sigma_i \otimes \sigma_j
    \right)\ ,
\end{equation}
where $x_i=\mathrm{tr}\left[\rho( \sigma_i \otimes I_2)\right]$, $y_i= \mathrm{tr}\left[\rho ( I_2\otimes \sigma_i )\right]$ are the Bloch vectors of the reduced states and $T_{ij}=\mathrm{tr}(\rho \sigma_i \otimes \sigma_ j)$ are the entries of its correlation matrix.

Concurrence is a measure of entanglement defined as \cite{Woo:98}
\begin{equation}
    \mathcal{C}(\rho)=\max (0,\lambda_1-\lambda_2-\lambda_3-\lambda_4)\ ,
\end{equation}
where $\lambda_i$ are the eigenvalues, in decreasing order, of the matrix
$M=\sqrt{\sqrt{\rho}\mu(\rho)\sqrt{\rho}}$, being $\mu(\rho):=(\sigma_y \otimes \sigma_y)\rho^{*}(\sigma_y \otimes \sigma_y)$. In the case of pure states, it reduces to the simpler expression $\mathcal{C}(\rho)=\sqrt{2(1-|x|^2)}$.
Negativity, denoted by $\mathcal{N}(\rho)$, is defined in terms of the trace norm of the partial transpose of the density matrix \cite{Vid.Wer:02}, namely,
\begin{equation}
\mathcal{N}(\rho)=\frac{|\rho^{T_A}|_1-1}{2}\,
\end{equation}
where $|X|_1:=\mathrm{tr}\sqrt{X^{\dagger}X}$ and $\rho^{T_A}$ denotes the partial transpose of the density matrix $\rho$ with respect to the subsystem $A$.
The von Neumann entropy 
\begin{equation}
    S(\rho)=-\sum_i \lambda_i \log \lambda_i\ ,
\end{equation}
captures how mixed a quantum state is,
where $\lambda_i$ are the eigenvalues of $\rho$. 
Another interesting quantity based on the purity of the state is the linear entropy $S_{\text{L}}$ \cite{Bag.Mar:17} (also called \textit{$1$-anticoherence}) 
   \begin{equation}
        S_{\text{L}}(\rho)=2(1-\nu_1^2-\nu_2^2)=2(1-\text{Tr}(\rho^2)) ,
    \end{equation}
where $\nu_i$, $i=1,2$, are the eigenvalues of the reduced density matrix traced on either subspace. Like entropy, linear entropy measures the \textit{mixedness} (or equivalently, the purity) of the quantum system \cite{Wei.Nemoto.Goldbart:2003}.
    Geometric quantum discord \cite{Dak.Ved.Bru:10} is a measure of quantum correlations related to quantum discord, but easier to compute, both capturing nonclassical properties not necessarily encoded in quantum entanglement. The geometric discord of a state $\rho$ of the above form (\ref{state2q}) is given by~\cite{Dak.Ved.Bru:10}
    \begin{equation}
    D_G(\rho)
    =
    \frac{1}{4}
    \left(
    ||y y^T||_2 + ||T||_2^2-k   \right) 
    \ ,
\end{equation}
where $k$ is the largest eigenvalue of the matrix $yy^T+T^TT$. We will work with $\mathcal{D}_G:=2D_G$ in our analysis, instead of the geometric quantum discord $D_g$, to have a quantity normalized to one.

Due to the complexity of the analytical expressions for the squared norm of acceleration and velocity when using the Bures metric, we decided to perform a numerical analysis using Mathematica. We initiated the process by randomly generating pure spin-$1$ states (hermitian matrices of dimension $3$, with a trace equal to $1$, and satisfying the condition $\rho^2=\rho$). Each one of these states corresponds to the symmetric part of a bipartite state composed of two qubits. Subsequently, we created $3000$ mixtures of these pure states with random weights, \ie, $3000$ random spin-$1$ mixed symmetric states.

In the following plots, where each dot represents a randomly generated state, we compare the above presented quantities with averaged speed and acceleration, at both levels, of the full system and of the subsystems. 
In plots \ref{fig:acc1_SR},
\ref{fig:acc1_VR},   and \ref{fig:VR_SLR} below, the numerical exploration suggests a  direct functional relation between the two compared variables. In contrast, in the rest of the plots,  a ``cloud'' of points appears, rather than a well-defined curve --- in those cases, it would be of interest to try and identify the boundary curves delimiting the ``clouds''.  
\begin{figure}
  \centering
  \begin{subfigure}{0.3\textwidth}
    \centering
\includegraphics[width=\linewidth]{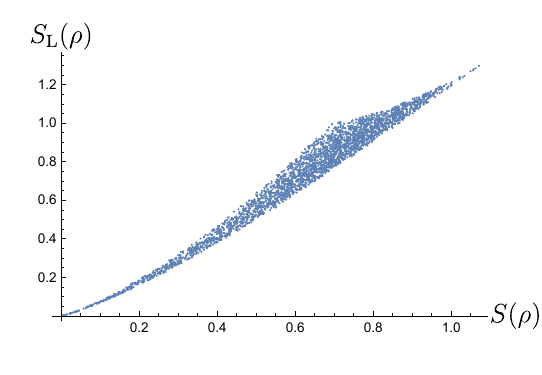}
\caption{$S_\text{L}(\rho)$ \emph{vs} $S(\rho)$.}
        \label{fig:S_SL}
  \end{subfigure}
  \begin{subfigure}{0.3\textwidth}
    \centering
\includegraphics[width=\linewidth]{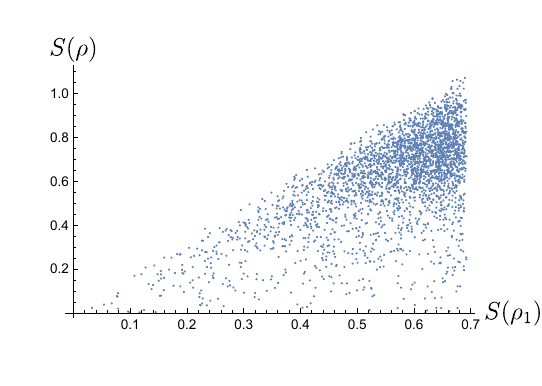}
\caption{$S(\rho)$ \emph{vs} $S(\rho_1)$.}
      \label{fig:SR_S}
  \end{subfigure}
  \begin{subfigure}{0.3\textwidth}
    \centering
\includegraphics[width=\linewidth]{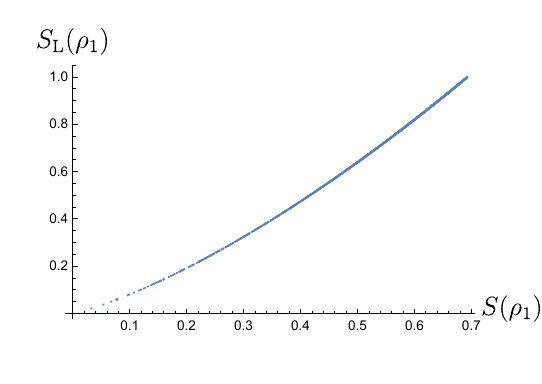}
\caption{$S_{\text{L}}(\rho_1)$ \emph{vs} $S(\rho_1)$.}
      \end{subfigure}
  \caption{Comparison between purity measures. Plots (a) and (c) show the linear entropy versus von Neumann entropy of the full state, and of the reduced density matrix, respectively. Notice that in the latter case, the relation is functional since both are measures of purity, in the case of (a) only for extreme  values of purity the relation tends to a bijective one. The von Neumann entropy of the full system and of the reduced state are compared in (b), where no functional relation is found, nonetheless, the numerical exploration suggests that there exists an upper bound: the purity of $\rho$ is constrained by the purity of the reduced states, $S(\rho_1)\geq \lambda S(\rho)$.}
\end{figure}
\begin{figure}
  \centering
  \begin{subfigure}{0.3\textwidth}
    \centering
\includegraphics[width=\linewidth]{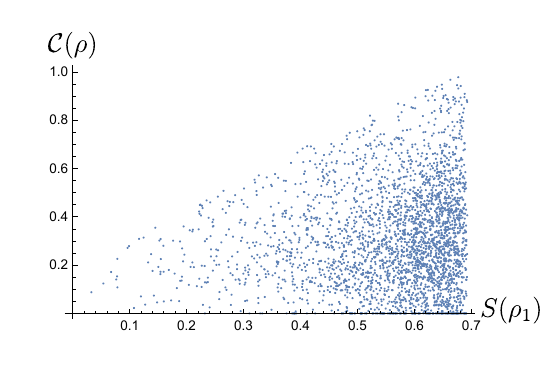}
\caption{$\mathcal{C}(\rho)$ \emph{vs} $S(\rho_1)$.}
        \label{fig:SR_C}
  \end{subfigure}
  \begin{subfigure}{0.3\textwidth}
    \centering
\includegraphics[width=\linewidth]{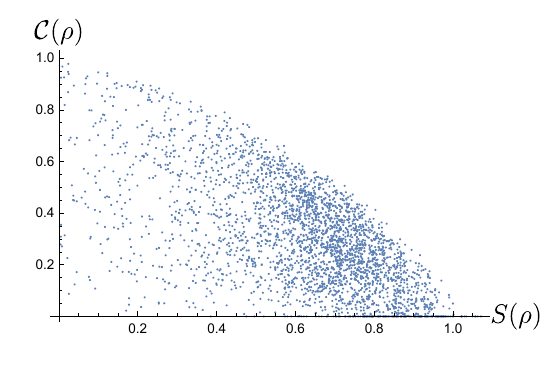}
\caption{$\mathcal{C}(\rho)$ \emph{vs} $S(\rho)$.}
      \label{fig:SR_T}
  \end{subfigure}
  \begin{subfigure}{0.3\textwidth}
    \centering
\includegraphics[width=\linewidth]{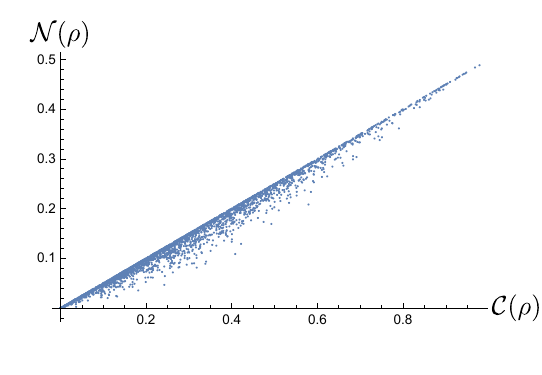}
\caption{$\mathcal{N}(\rho)$ \emph{vs} $\mathcal{C}(\rho)$.}
      \end{subfigure}
  \caption{Comparison between purity measures and entanglement measures. Plots (a) and (b) show the concurrence versus von Neumann entropy of the reduced density matrix, and the full density matrix, respectively. Plot  (a) seems to suggest a relation of the form  $S(\rho_1)\geq \lambda \mathcal{C}(\rho)+\mu$, valid for $S(\rho_1) \geq .2$. In (c) two measures of entanglement, negativity and concurrence, are compared, showing an approximate proportionality relation, valid for most of the states.}
\end{figure}
\begin{figure}
  \centering
  \begin{subfigure}{0.28\textwidth}
    \centering
\includegraphics[width=\linewidth]{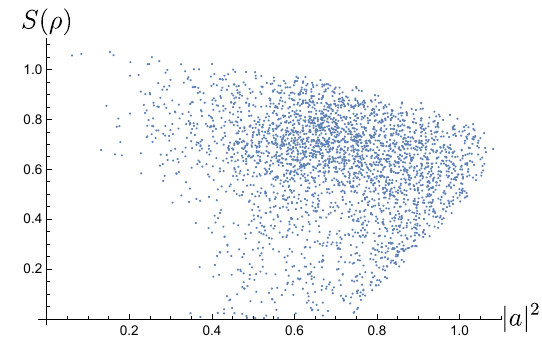}
    \caption{$\mathcal{S}(\rho)$ \emph{vs} $|a|^2$.}
    \label{fig:acc_S}
  \end{subfigure}
  \begin{subfigure}{0.28\textwidth}
    \centering
 \includegraphics[width=\linewidth]{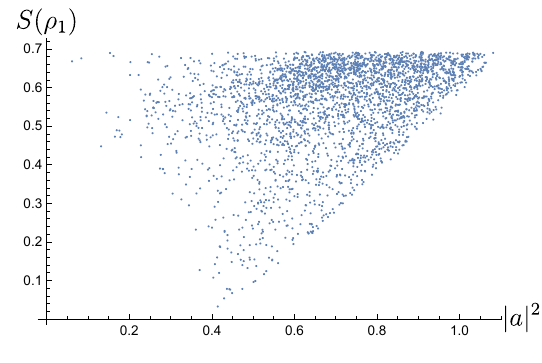}
    \caption{$\mathcal{S}(\rho_1)$ \emph{vs} $|a|^2$.}
    \label{fig:acc_SR}
  \end{subfigure}
  \begin{subfigure}{0.28\textwidth}
    \centering
    \includegraphics[width=\linewidth]{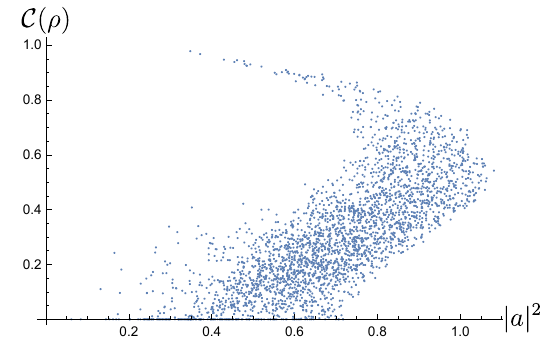}
    \caption{$\mathcal{C}(\rho)$ \emph{vs} $|a|^2$.}
    \label{fig:acc_C}
  \end{subfigure}
\begin{subfigure}{0.28\textwidth}
    \centering
    \includegraphics[width=\linewidth]{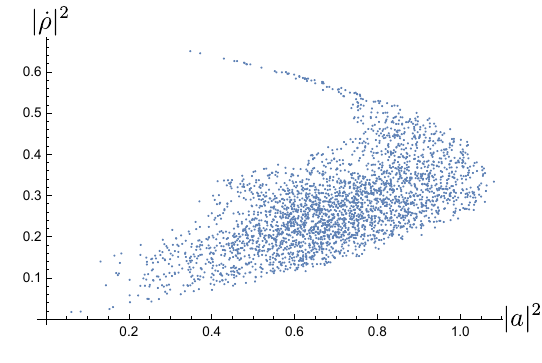}
    \caption{$|\dot{\rho}|^2$ \emph{vs} $|a|^2$.}
     \end{subfigure}
  \begin{subfigure}{0.28\textwidth}
    \centering
    \includegraphics[width=\linewidth]{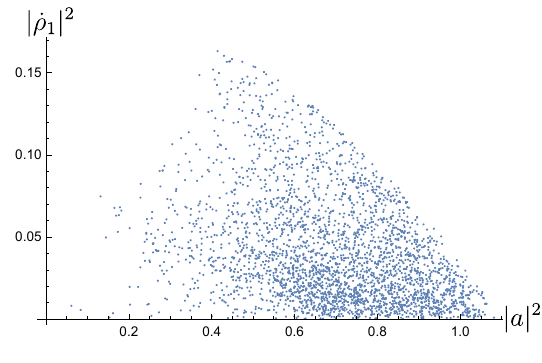}
    \caption{$|\dot{\rho}_1|^2$ \emph{vs} $|a|^2$.}
    \label{fig:acc_VR}
  \end{subfigure}
  \begin{subfigure}{0.28\textwidth}
    \centering
    \includegraphics[width=\linewidth]{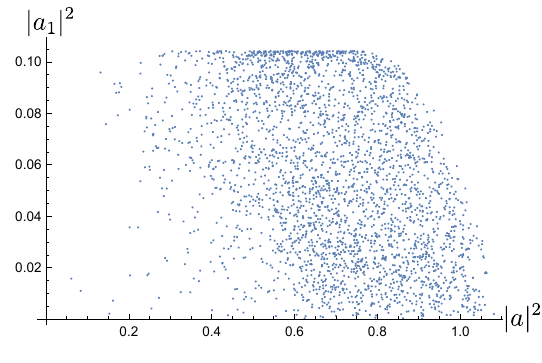}
    \caption{$|a_1|^2$ \emph{vs} $|a|^2$.}
   \end{subfigure}
\caption{Comparison between the squared norm of the acceleration $|a|^2$ and other quantities. In subfigures (a) and (c), we can identify two boundaries enclosing the region. In (c) we can also note the definition of potential borders. Note in Figure (b) that for $|a|^2 > 0.4$, the measure of purity should exceed certain values defined by the boundary of the region.}
  \end{figure}
\begin{figure}
  \centering
 \begin{subfigure}{0.3\textwidth}
    \centering
    \includegraphics[width=\linewidth]{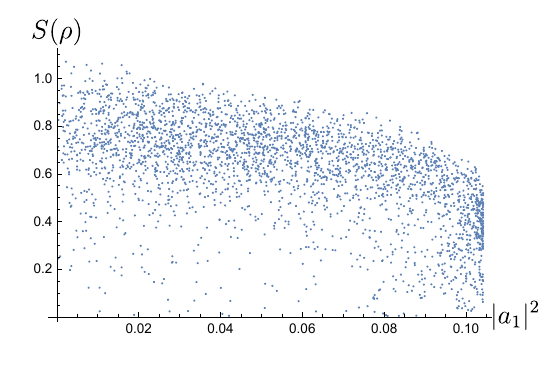}
    \caption{$\mathcal{S}(\rho)$ \emph{vs} $|a_1|^2$.}
    \label{fig:acc1_S}
  \end{subfigure}
  \begin{subfigure}{0.3\textwidth}
    \centering
    \includegraphics[width=\linewidth]{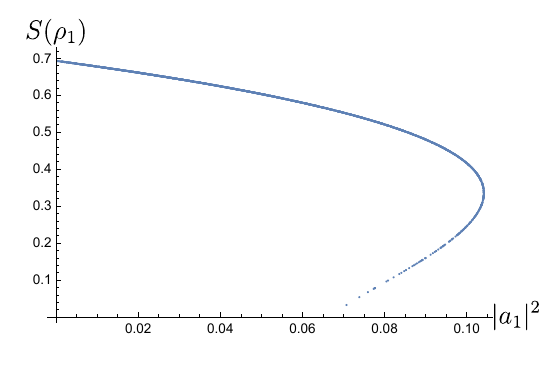}
    \caption{ $\mathcal{S}(\rho_1)$ \emph{vs} $|a_1|^2$.}
    \label{fig:acc1_SR}
  \end{subfigure}
  \begin{subfigure}{0.3\textwidth}
    \centering
    \includegraphics[width=\linewidth]{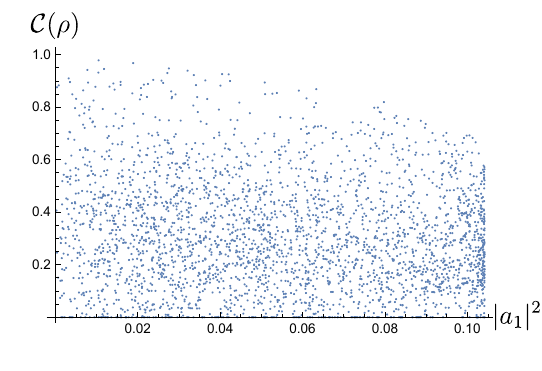}
    \caption{$\mathcal{C}(\rho)$ \emph{vs} $|a_1|^2$.}
    \label{fig:acc1_C.pdf}
  \end{subfigure}
  \begin{subfigure}{0.3\textwidth}
    \centering
\includegraphics[width=\linewidth]{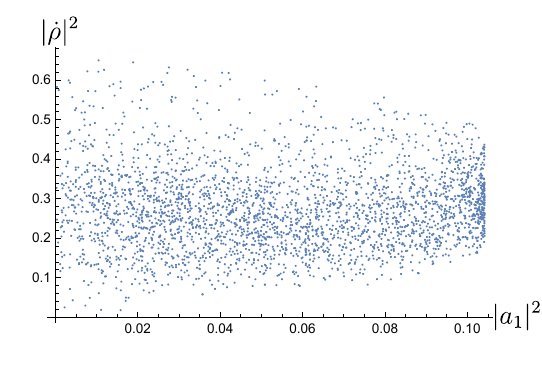}
    \caption{$|\dot{\rho}|^2$ \emph{vs} $|a_1|^2$.}
    \label{fig:acc1_V.pdf}
  \end{subfigure}
  \begin{subfigure}{0.3\textwidth}
    \centering
\includegraphics[width=\linewidth]{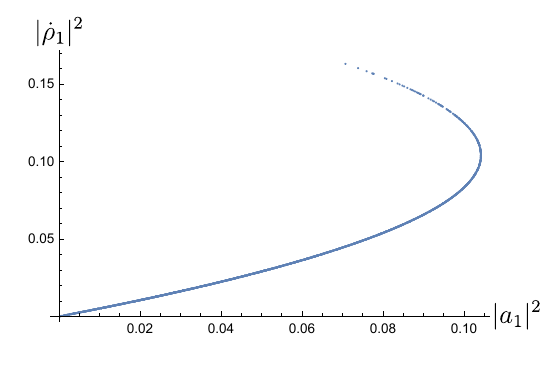}
    \caption{$|\dot{\rho}_1|^2$ \emph{vs} $|a_1|^2$.}
    \label{fig:acc1_VR}
  \end{subfigure}
  \caption{Comparison between the squared norm of the acceleration $|a_1|^2$ (corresponding to the reduced density matrices) and other quantities. In subfigures (b),  (e)  we note a clear functional relationship between the quantities involved.}
\end{figure}
\begin{figure}[H]
  \centering
  \begin{subfigure}{0.28\textwidth}
    \centering
\includegraphics[width=\linewidth]{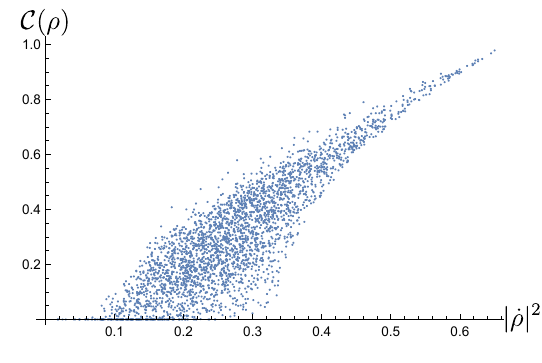}
    \caption{$\mathcal{C}(\rho)$ \emph{vs} $|\dot{\rho}|^2$. }
    \label{fig:V_C}
  \end{subfigure}
  \begin{subfigure}{0.28\textwidth}
    \centering
\includegraphics[width=\linewidth]{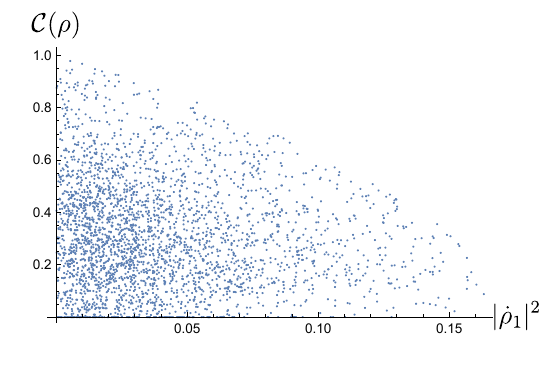}
    \caption{$\mathcal{C}(\rho)$ \emph{vs} $|\dot{\rho}_1|^2$.}
    \label{fig:VR_C}
  \end{subfigure}
  \begin{subfigure}{0.28\textwidth}
    \centering
    \includegraphics[width=\linewidth]{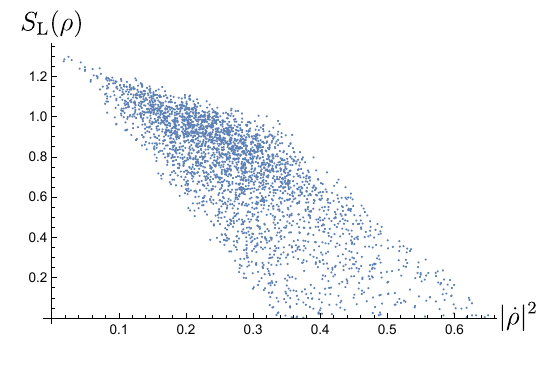}
    \caption{$\mathcal{S}_{\text{L}}(\rho)$ \emph{vs} $|\dot{\rho}|^2$.}
      \end{subfigure}
 \begin{subfigure}{0.28\textwidth}
    \centering
    \includegraphics[width=\linewidth]{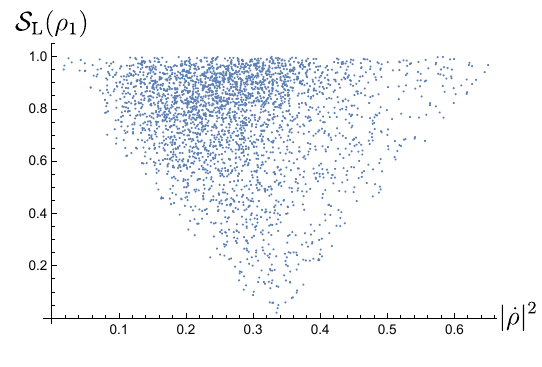}
    \caption{ $\mathcal{S}_{\text{L}}(\rho_1)$ \emph{vs} $|\dot{\rho}|^2$.}
    \label{fig:V_SLR}
     \end{subfigure}
  \begin{subfigure}{0.28\textwidth}
    \centering
\includegraphics[width=\linewidth]{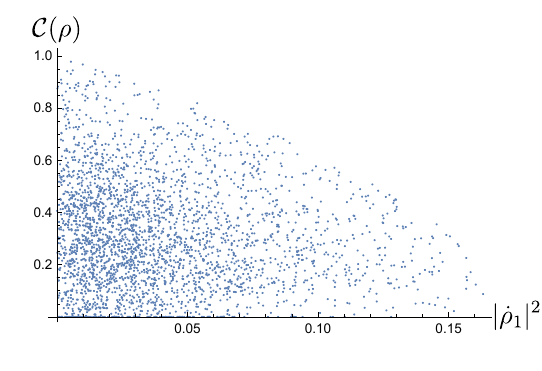}
    \caption{ $\mathcal{C}(\rho)$ \emph{vs} $|\dot{\rho}_1|^2$.}
     \end{subfigure}
  \begin{subfigure}{0.28\textwidth}
    \centering
\includegraphics[width=\linewidth]{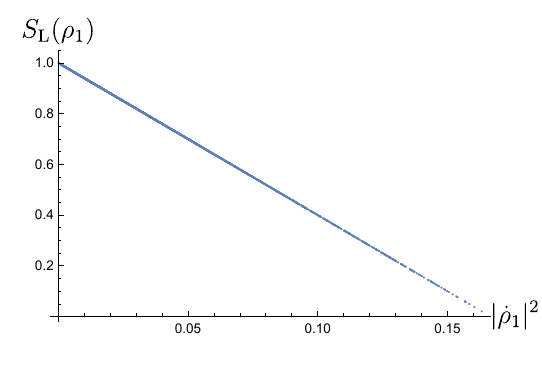}
    \caption{ $\mathcal{S}_{\text{L}}(\rho_1)$ \emph{vs} $|\dot{\rho}_1|^2$.}
    \label{fig:VR_SLR}
   \end{subfigure}
  \caption{Comparison between the squared norm of the velocities $|\dot{\rho}|^2$, $|\dot{\rho}_1|^2$ and other quantities. Note in subfigure (a) that, apparently, for each degree of entanglement, there exists an allowable region of $|\dot{\rho}|^2$, and for high degrees of entanglement ($\mathcal{C}>0.6$), this region becomes narrower. We observe that maximally entangled states also exhibit the highest value of $|\dot{\rho}|^2$. In subfigures (b)-(e) we can note that it seems that there are definite regions for the possible simultaneous values of both variables. In subfigure (f), we identify a linear relationship between the purity and the speed of the reduced states.}
  \end{figure}
\begin{figure}[H]
  \centering
 \begin{subfigure}{0.3\textwidth}
    \centering
\includegraphics[width=\linewidth]{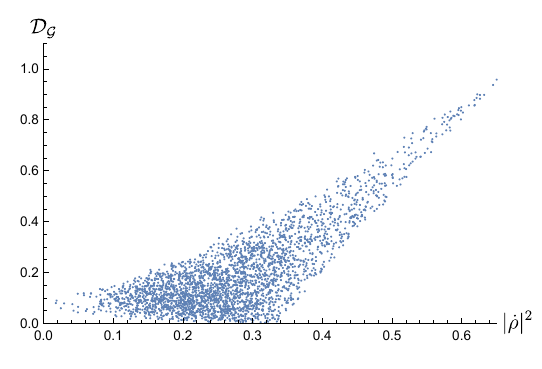}
    \caption{$\mathcal{D}_{\mathcal{G}}$ \emph{vs} $|\dot{\rho}|^2$.}
    \label{fig:V_D}
  \end{subfigure}
  \begin{subfigure}{0.3\textwidth}
    \centering
\includegraphics[width=\linewidth]{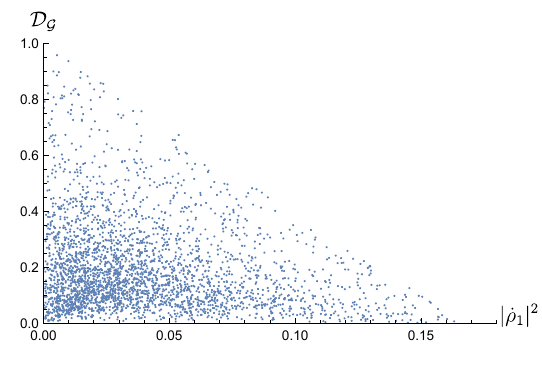}
    \caption{$\mathcal{D}_{\mathcal{G}}$ \emph{vs} $|\dot{\rho}_1|^2$.}
    \label{fig:VR_D}
  \end{subfigure}
  \begin{subfigure}{0.3\textwidth}
    \centering
\includegraphics[width=\linewidth]{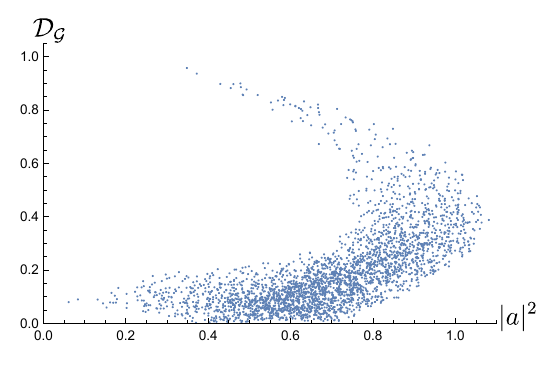}
    \caption{$\mathcal{D}_{\mathcal{G}}$ \emph{vs} $|a|^2$ }
  \end{subfigure}
  \begin{subfigure}{0.3\textwidth}
    \centering
\includegraphics[width=\linewidth]{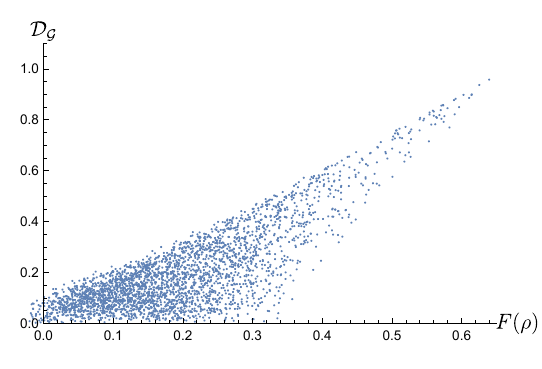}
    \caption{$\mathcal{D}_{\mathcal{G}}$ \emph{vs} $F(\rho)$ }
  \end{subfigure}
  \begin{subfigure}{0.3\textwidth}
    \centering
\includegraphics[width=\linewidth]{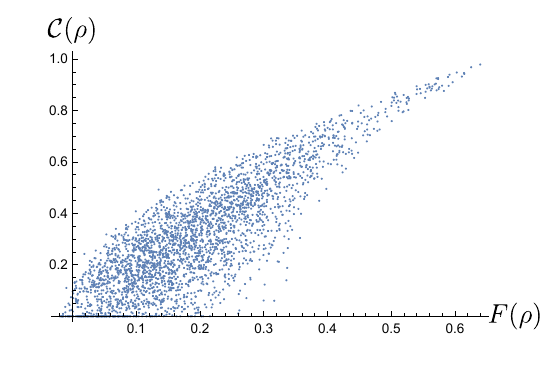}
    \caption{$\mathcal{C}(\rho)$ \emph{vs} $F(\rho)$. }
      \end{subfigure}
  \caption{Comparison between kinematical quantities and geometric quantum discord $\mathcal{D}_{\mathcal{G}}$ is shown in subfigures (a)-(d). Subfigure (e) compares the speed excess with the concurrence --- note that the relationship between them is not linear, unlike in the case of pure states. Note also that in (d) and (e) a very small percentage of states has negative $F$.}
  \label{kinD_Fig}
\end{figure}
\section{Concluding remarks}
\label{Conclusions}
We summarize the main points of this work:
(i) The total variance of a pure spin state is a measure of its squared rotational speed, averaged over all rotation axes. The concept is generalized for mixed states using the Bures metric. 
(ii)
Entanglement increases rotational speed, hence the relevance of total variance as an entanglement  measure. The addition law for total variance is pythagorean for pure separable states, $|v|^2=|v_1|^2+|v_2|^2$, and receives additional positive contributions for pure entangled states. Speed excess, defined as $|v|^2-|v_1|^2-|v_2|^2$,  may thus be used to quantify entanglement. 
(iii)
Total (average, squared, rotational) acceleration may be similarly defined for both pure and mixed states. For separable states we find $|a|^2=|a_1|^2+|a_2|^2+4 |v_1|^2 |v_2|^2$, which, incidentally, means that the quantity $|a|^2-2|v|^4$ is additive under system composition.  We found a simple analytical formula for the acceleration of a mixed state according to the Bures metric  (Eq.~(\ref{eq:accelerationBures})). Numerical results, exploring the correlation of total acceleration with other physical characteristics of the state, display the full gamut of possibilities. As shown in (d), (e) of Fig.~\ref{kinD_Fig}, speed excess can be negative for a very small percentage of state space volume. 

Some directions for further work along similar lines include:
(i)
The additivity of $|a|^2-2|v|^4$ under separable system composition suggests exploring the physical significance of this quantity. 
(ii)
The correlation of total acceleration with other relevant physical quantities, which we started  in Sect.~\ref{KQPaQC}, should also be pursued analytically. 
(iii) 
Higher-order Bures metric covariant derivative formulas like~(\ref{eq:accelerationBures}) would be desirable. The physical significance of the average of the modulus squared of these quantities should be explored. Do the states that extremize these quantities have any desirable physical properties?
(iv)
What are the physical characteristics of the states that have negative speed excess?

\section*{Acknowledgements}
ESE acknowledges support from the postdoctoral fellowship of the IPD-STEMA program of the University of Liège (Belgium).
\appendix
\section{Calculating the average of the norm squared of the acceleration}
\label{AtHd}
%
The average norm of the acceleration is given by
\begin{equation}
\langle |\dot{v}|^2 \rangle = 
\left( \langle T_A T_B T_C T_D \rangle - 4\langle T_A T_B T_C \rangle \langle T_D \rangle - 
\langle T_A T_B \rangle \langle T_C T_D \rangle  +8  
\langle T_A T_B \rangle \langle T_C \rangle \langle T_D \rangle - 4 
\langle T_A \rangle \langle T_B \rangle \langle T_C \rangle \langle T_D \rangle \right)
\int_{S^2} H^A H^B H^C H^D \, ,
\end{equation}
where $H=H^A T_A$. The previous equation can be written as the linear combination of the angular momentum operators, or as the linear combination of the tensor operators $\{ T_{1m} \}_{m=-1}^1$
\begin{equation}
H= \sum_{\alpha=x,y,z} n_{\alpha}S_{\alpha} = \sum_{m=-1}^1 r_m T_{1m}^{(s)} \, .
\end{equation}
with
\begin{align}
n_x = & \frac{A}{\sqrt{2}} \left(  
r_{-1} -r_1 \right) \, \quad n_y = -i \frac{A}{\sqrt{2}} \left( r_{-1} + r_1  \right) \, , \quad n_z = A r_0 \, , 
\\
r_1 = & -\frac{1}{\sqrt{2}A}\left( n_x - i n_y \right) \, , \quad 
r_{-1} = \frac{1}{\sqrt{2}A}\left( n_x + i n_y \right) \, , \quad 
r_0 = \frac{n_z}{A} \, ,
\end{align}
where $A=A(s) = \sqrt{\frac{3}{s(s+1)(2s+1)}}$. The average integrals of the product of components $r_{m}$ are equal to
\begin{equation}
\label{int.M}
\int_{S^2}r_0^4 = \frac{1}{5A^4} \, , \quad 
\int_{S^2} r_0^2 r_1 r_{-1} = -\frac{1}{15A^4} \, ,
\quad
\int_{S^2} r_1^2 r_{-1}^2 = \frac{2}{15A^4} \, ,
\end{equation}
and the rest are equal to zero.
As a concrete application of the above scheme, consider the term $\int_{S^2} \langle H^A H^B H^C H^D T_A T_B T_C T_D \rangle$, expressed in terms of Wigner-D matrices $D^{(s)}_{mm'}=D^{(s)}_{mm'}(\alpha, \beta, \gamma)$
\begin{align}
\int_{S^2} \langle  H^4 \rangle = r_0^4 \int_{S^2} \langle (D^{(s)} T_{10} D^{(s) \dagger})^4 \rangle = A^{-4} \int_{S^2} \langle D^{(s)} T_{10}^4 D^{(s) \dagger} \rangle
\end{align}
taking $n_z=1$. Now, we make use that the multipolar operators is a basis. In particular, we know that
\begin{equation}
T_{l_1m_1} T_{l_2m_2} = (-1)^{2l_2+l-2s}\sqrt{(2l_1+1)(2l_2+1)} \sjs{l_1}{l_2}{l}{s}{s}{s} c_{l_1m_1,l_2m_2}^{lm} T_{lm} \, \equiv \chi(l_1,l_2,l;s) c_{l_1m_1,l_2m_2}^{lm} T_{lm} \, ,
\end{equation}
where we use the Einstein convention. Hence, we can calculate the expansion of $T_{10}^4$ in the multipolar tensor basis.
\begin{equation}
T_{10}^4 = \chi(1,1,l;s) \chi(1,1,l';s) c_{10,10}^{lm} c_{10,10}^{l'm'} T_{lm} T_{l'm'} = \chi(1,1,l;s) \chi(1,1,l';s) \chi(l,l',L;s) c_{10,10}^{l0} c_{10,10}^{l'0} c_{l0,l'0}^{LM} T_{LM} \, .
\end{equation}
Now, we use the result that
\begin{equation}
\int_{S^2} D^{(s)} T_{LM} D^{(s) \dagger} = \int_{S^2} D^{(L)}_{M' M}(\phi,\theta,0) T_{L M'} = \delta_{M0} \delta_{L0} T_{00} \, .
\end{equation}
Hence,
\begin{align}
\int_{S^2} \langle  H^4 \rangle = & A^4 \chi(1,1,l;s) \chi(1,1,l';s) \chi(l,l',0;s) c_{10,10}^{l0} c_{10,10}^{l'0} c_{l0,l'0}^{00} \langle T_{00}  \rangle
\nonumber
\\
= & \sum_{L=0}^2 \frac{A^{-4}}{\sqrt{2s+1}}\chi(1,1,L;s)^2  \chi(L,L,0;s) c_{10,10}^{L0} c_{10,10}^{L0} c_{L0,L0}^{00} \, .
\label{ans.1}
\end{align}
In the same way, we calculate the next two terms
\begin{align}
\langle D^{(s)} T_{10} T_{10}T_{10} D^{(s) \dagger} \rangle \langle D^{(s)} T_{10} D^{(s) \dagger} \rangle = &
\chi(1,1,l;s) c_{10,10}^{l0} D^{(1)}_{M0}
\langle D^{(s)} T_{l0} T_{10} D^{(s) \dagger} \rangle \langle  T_{1M}  \rangle
\\
= &
\chi(1,1,l;s) \chi(l,1,L;s) c_{10,10}^{l0} c_{l0,10}^{L0} D^{(1)}_{M0}
\langle D^{(s)} T_{L0} D^{(s) \dagger} \rangle \langle  T_{1M}  \rangle
\\
= &
\chi(1,1,l;s) \chi(l,1,L;s) c_{10,10}^{l0} c_{l0,10}^{L0} D^{(1)}_{M0} D^{(L)}_{M' 0}
\langle  T_{L M'} \rangle \langle  T_{1M}  \rangle \, .
\end{align}
We now use the integral over two Wigner-D matrices
\begin{equation}
D^{(L)}_{M N} = (-1)^{N-M} D^{(L) \dagger}_{-M -N} \, ,
\end{equation}
\begin{equation}
\int_{S^2} D^{(L)}_{M N}(\phi,\theta, 0) D^{(L')}_{M' N'}(\phi,\theta, 0) = (-1)^{N'-M'} \int_{S^2} D^{(L)}_{M N} D^{(L') \dagger}_{-M' -N'} = 
 \frac{(-1)^{M-N}}{2L+1} \delta_{L L'} \delta_{M-M'} \delta_{N-N'} \, .
\end{equation}
Then, we obtain that
\begin{align}
\int_{S^2} \langle D^{(s)} T_{10} T_{10}T_{10} D^{(s) \dagger} \rangle \langle D^{(s)} T_{10} D^{(s) \dagger} \rangle
= & 
\frac{1}{2L+1}\chi(1,1,l;s) \chi(l,1,L;s) c_{10,10}^{l0} c_{l0,10}^{L0} (-1)^{M} \delta_{L1} \delta_{M-M'}
\langle  T_{L M'} \rangle \langle  T_{1M}  \rangle
\nonumber
\\
= & 
\frac{1}{3} \chi(1,1,l;s) \chi(l,1,1;s) c_{10,10}^{l0} c_{l0,10}^{10} (-1)^{M}  \langle  T_{1 -M} \rangle \langle  T_{1M}  \rangle
\nonumber
\\
= & 
\frac{1}{3} \chi(1,1,l;s) \chi(l,1,1;s) c_{10,10}^{l0} c_{l0,10}^{10} \langle  T_{1 M}^{\dagger} \rangle \langle  T_{1M}  \rangle \, .
\end{align}
Therefore
\begin{equation}
\int_{S^2} \langle H^3 \rangle \langle H \rangle = \frac{1}{3A^{4}} \chi(1,1,L;s) \chi(L,1,1;s) c_{10,10}^{L0} c_{L0,10}^{10} \langle  T_{1 M}^{\dagger} \rangle \langle  T_{1M}  \rangle \, .
\label{ans.2}
\end{equation}
Similarly, we calculate
\begin{equation}
\int_{S^2}\langle H^2 \rangle \langle H^2 \rangle = \frac{\left[ \chi(1,1,L;s) c_{10,10}^{L0} \right]^2}{A^4(2L+1)} 
\langle  T_{L M}^{\dagger} \rangle \langle  T_{LM}  \rangle
\label{ans.3}
\end{equation}
For the next term, we use the following formula (Eq. 4, p.96 of~\cite{Var.Mos.Khe:88})
\begin{equation}
\int_{S^2} D_{M_3 M_3'}^{(J_3)} D_{M_2 M_2'}^{(J_2)} D_{M_1 M_1'}^{(J_1)}
= \frac{(-1)^{M_3-M_3'}}{2J_3+1}
c_{J_1 M_1 J_2 M_2}^{J_3-M_3}
c_{J_1 M_1' J_2 M_2'}^{J_3-M_3'} \, ,
\end{equation} 
\begin{align}
\int_{S^2} \langle H^2 \rangle
 \langle H \rangle^2
  = & A^{-4} \int_{S^2} 
  \langle D T_{10} T_{10} D^{\dagger} \rangle
  \langle D T_{10} D^{\dagger} \rangle
  \langle D T_{10} D^{\dagger} \rangle
\nonumber
\\
  = & A^{-4} c_{1010}^{L0} \chi(1,1,L;s)
  \langle T_{LM} \rangle
  \langle T_{1 N_1} \rangle
  \langle T_{1 N_2} \rangle
  \int_{S^2} D_{M0}^{(L)} D_{N_10}^{(1)}D_{N_20}^{(1)} 
\nonumber
\\
  = & A^{-4} \langle T_{LM} \rangle
  \langle T_{1 N_1} \rangle
  \langle T_{1 N_2} \rangle
\frac{(-1)^{M}}{2L+1}
c_{1 N_2 1 N_1}^{L-M}
c_{1 0 1 0}^{L 0}
c_{1 0 1 0}^{L 0} \chi(1,1,L;s)
\nonumber
\\
  = &
\frac{c_{1 N_2 1 N_1}^{LM}
(c_{1 0 1 0}^{L 0})^2 }{A^4(2L+1)}  
\chi(1,1,L;s)
  \langle T_{LM}^{\dagger} \rangle
  \langle T_{1 N_1} \rangle
  \langle T_{1 N_2} \rangle
\label{ans.4}
\end{align}
Finally, we calculate the last term with the integrals of Eq. \eqref{int.M}
\begin{align}
\int_{S^2} \langle H \rangle^4 = & \frac{1}{15A^4} \left[ 
3 \langle T_{10} \rangle^4 - 12 \langle T_{10} \rangle^2 \langle T_{11} \rangle \langle T_{1-1} \rangle + 12
\langle T_{11} \rangle^2
\langle T_{1-1} \rangle^2
\right]
\nonumber
\\
=& 
\frac{1}{5A^4} \left[ 
\langle T_{10} \rangle^2
-2
\langle T_{11} \rangle
\langle T_{1-1} \rangle
\right]^2
\nonumber
\\
= &
\frac{1}{5A^4} \left[ 
\langle T_{1M} \rangle
\langle T_{1M}^{\dagger} \rangle
\right]^2
\label{ans.5}
\end{align}
The final result is given by Eqs. \eqref{ans.1}, \eqref{ans.2}, \eqref{ans.3}, \eqref{ans.4} and \eqref{ans.5},
\begin{align}
\langle |\dot{v}|^2 \rangle = &  
A^{-4} \Big( \sum_{L=0}^2 \Big\{ \frac{1}{\sqrt{2s+1}} \big[ \chi(1,1,L;s) c_{10,10}^{L0} \big]^2  \chi(L,L,0;s)  c_{L0,L0}^{00}
-\frac{4}{3} \chi(1,1,L;s) \chi(L,1,1;s) c_{10,10}^{L0} c_{L0,10}^{10} \langle  T_{1 M}^{\dagger} \rangle \langle  T_{1M}  \rangle
\nonumber
\\
&
-
\frac{\left[ \chi(1,1,L;s) c_{10,10}^{L0} \right]^2 }{2L+1}
\langle  T_{L M}^{\dagger} \rangle \langle  T_{LM}  \rangle
+ 
\frac{8 \chi(1,1,L;s) c_{1 N_2 1 N_1}^{LM}
\big[c_{1 0 1 0}^{L 0} \big]^2
}{2L+1}
\langle T_{LM}^{\dagger} \rangle
  \langle T_{1 N_1} \rangle
  \langle T_{1 N_2} \rangle \Big\}
  \\
  &
 -\frac{4}{5} \left[ 
\langle T_{1M} \rangle
\langle T_{1M}^{\dagger} \rangle
\right]^2 \Big) \, .
\label{Fin.1}
\end{align}
Some quantities in the previous equations can be simplified by using formulas for the 6j-symbols and the CG coefficients. For example, some quantities of Eq.~\eqref{Fin.1} are simplified to
\begin{equation}
\label{ANS.1}
A^{-4} \sum_{L=0}^2 \frac{1}{\sqrt{2s+1}} \big[ \chi(1,1,L;s) c_{10,10}^{L0} \big]^2  \chi(L,L,0;s)  c_{L0,L0}^{00}
= 
\left\{ 
\begin{array}{c c}
\frac{s^2(s+1)^2}{9}  & s \leq 1/2
\\
\frac{s(s+1)(3s^2+3s-1)}{15} & s \geq 1
\end{array}
\right. \, .
\end{equation}
\begin{equation}
\label{ANS.2}
\frac{4}{3} A^{-4} \sum_{L=0}^2 
\chi(1,1,L;s) \chi(L,1,1;s) c_{10,10}^{L0} c_{L0,10}^{10} 
= 
\left\{
\begin{array}{c c}
\frac{4}{27}s^2(s+1)^2(2s+1) & s \leq 1/2
\\
\frac{4}{45}s(s+1)(2s+1)(3s^2+3s-1) & s \geq 1
\end{array}
\right. \, .
\end{equation}
\begin{equation}
\label{ANS.3}
\frac{\left[ \chi(1,1,L;s) c_{10,10}^{L0} \right]^2 }{A^4(2L+1)}
=
\left\{
\begin{array}{c c}
\frac{s^2(s+1)^2(2s+1)}{9} & L=0
\\
0
& L=1
\\
\frac{s(s+1)(2s-1)(2s+1)(2s+3)}{225}
& L=2
\end{array}
\right.
\, .
\end{equation}
\begin{equation}
\label{ANS.4}
\frac{8 \chi(1,1,L;s) 
\big[c_{1 0 1 0}^{L 0} \big]^2
}{A^4(2L+1)}
=
\left\{
\begin{array}{c c}
-\frac{8}{9\sqrt{3}}s^2(s+1)^2(2s+1)^{3/2}
 & L=0
 \\
 0 & L=1
 \\
\frac{8}{45\sqrt{30}} s(s+1)(2s+1)
\sqrt{(2s+3)(2s+2)(2s+1)2s(2s-1)}
  & L=2 
\end{array}
\right.
\, ,
\end{equation}
leading to Eq.~(\ref{Fin.ans}) in the main text.

\end{document}